\documentclass[fleqn,usenatbib]{mnras}

\usepackage{newtxtext,newtxmath}

\usepackage[T1]{fontenc}
\usepackage{ae,aecompl}

\usepackage{graphicx}	
\usepackage{amsmath}	
\usepackage{amssymb}	
\usepackage{tablefootnote}
\usepackage{pdflscape}	

\def\lya{{\rm\,Ly$\alpha$}}
\def\ha{{\rm\,H$\alpha$}}
\def\hb{{\rm\,H$\beta$}}

\def\nii{{\rm\,[N{\sc ii}]}}

\def\oiii{{\rm\,[O{\sc iii}]}}

\def\hi{{\rm\,H{\sc i}}}

\def\paa{{\rm\,Pa$\alpha$}}
\def\msun{{\rm M}$_{\odot}$}

\title[MDCS--II. Massive galaxies in protoclusters]
{MAHALO Deep Cluster Survey II. Characterizing massive forming galaxies in the Spiderweb protocluster at $z=2.2$}

\author[R. Shimakawa et al.]{
Rhythm Shimakawa,$^{1}$\thanks{rhythm@naoj.org}
Yusei Koyama,$^{1}$
Huub J. A. R\"ottgering,$^{2}$
\newauthor
Tadayuki Kodama,$^{3}$
Masao Hayashi,$^{4}$
Nina A. Hatch,$^{5}$
Helmut Dannerbauer,$^{6,7}$
\newauthor
Ichi Tanaka,$^{1}$
Ken-ichi Tadaki,$^{4}$
Tomoko L. Suzuki,$^{3}$
Nao Fukagawa,$^{8}$
\newauthor
Zheng Cai$^{9}$
and Jaron D. Kurk$^{10}$
\\
$^{1}$Subaru Telescope, National Astronomical Observatory of Japan, National Institutes of Natural Sciences, 650 North A'ohoku Place, Hilo, HI 96720, USA\\
$^{2}$Leiden Observatory, Leiden University, PO Box 9513, NL-2300 RA Leiden, the Netherlands\\
$^{3}$Astronomical Institute, Tohoku University, Aoba-ku, Sendai 980-8578, Japan\\
$^{4}$National Astronomical Observatory of Japan, Osawa, Mitaka, Tokyo 181-8588, Japan\\
$^{5}$School of Physics and Astronomy, University of Nottingham, University Park, Nottingham NG7 2RD, UK\\
$^{6}$Instituto de Astrof\'{i}sica de Canarias, E-38205 La Laguna, Tenerife, Spain\\
$^{7}$Universidad de La Laguna Dpto. Astrof\'{i}sica, E-38206 La Laguna, Tenerife, Spain\\
$^{8}$Department of Astronomical Science, SOKENDAI, Osawa, Mitaka, Tokyo 181-8588, Japan\\
$^{9}$UCO/Lick Observatory, University of California, 1156 High Street, Santa Cruz, CA 95064, USA\\
$^{10}$Max-Planck-Institut f\"ur extraterrestrische Physik, Giessenbachstra{\ss}e 1, 85748 Garching, Germany
}

\date{Accepted 2018 September 20. Received 2018 September 1; in original form 2018 July 12.}

\pubyear{2018}

\begin{document}
\label{firstpage}
\pagerange{\pageref{firstpage}--\pageref{lastpage}}
\maketitle

\begin{abstract}
This paper is the second in a series presenting the results of our deep H$\alpha$-line 
survey towards protoclusters at $z>2$, based on narrow-band imaging with the Subaru 
Telescope. This work investigates massive galaxies in a protocluster region associated 
with a radio galaxy (PKS~1138$-$262), the Spiderweb galaxy, at $z=2.2$. Our 0.5 mag 
deeper narrow-band imaging than previous surveys collects a total of 68 H$\alpha$ 
emitters (HAE). 17 out of the 68 are newly discovered protocluster members. First, a 
very high characteristic stellar mass of M$_\star^\ast=10^{11.73}$ M$_\odot$ is 
measured from a Schechter function fit to the mass distribution of HAEs. Together with 
the Chandra X-ray data, we find that four out of six massive HAEs (M$_\star>10^{11}$ 
M$_\odot$) show bright X-ray emission, suggesting that they host active galactic 
nuclei (AGNs). Their mass estimates, therefore, would be affected by 
the nuclear emission from AGNs.
Notably, the X-ray detected HAEs are likely positioned near the boundary between 
star-forming and quiescent populations in the rest-frame $UVJ$ plane. 
Moreover, our deep narrow-band data succeed in probing the bright H$\alpha$ 
(+[N{\sc ii}]) line nebula of the Spiderweb galaxy extending over $\sim100$ physical 
kpc. These results suggest that the massive galaxies in the Spiderweb protocluster are 
on the way to becoming the bright red sequence objects seen in local galaxy clusters, 
where AGNs might play an essential role in their quenching processes. Though a more 
statistical database is needed to build a general picture. 
\end{abstract}




\section{Introduction}\label{s1}

It is well-known that massive quiescent galaxies are more predominant in the centres 
of galaxy clusters relative to the general fields in the present day. This trend is 
often characterised by colours 
\citep{Vaucouleurs:1961,Visvanathan:1977,Butcher:1984,Bower:1992,Bower:1998,Terlevich:2001,Tanaka:2005,Kodama:2007,Mei:2009,Bamford:2009,Peng:2010,Muzzin:2012,Wetzel:2012,Darvish:2016} 
and morphological types 
\citep{Dressler:1980,Dressler:1997,Couch:1998,Goto:2003,Kauffmann:2004,Wel:2008,Cappellari:2011,Houghton:2013,Fogarty:2014,Brough:2017,Lopes:2017}. 
Over ten billion years ago, the most massive structures in the Universe -- galaxy 
protoclusters -- played a prominent role in the star formation and mass assembly 
of massive galaxies \citep{Chiang:2017}. Massive protoclusters\footnotemark[1] 
\citep{Albada:1961,Peebles:1970a,Sunyaev:1972} at redshift $z\sim$ 2--3 are ideal 
test-beds to probe this rapid transition, and thus develop our understanding of which 
physical phenomena have driven such early and/or fast growth in centres of distant 
galaxy clusters 
\citep{Steidel:2005,Doherty:2010,Tanaka:2010b,Hatch:2011,Gobat:2011,Koyama:2013b,Tanaka:2013,Kubo:2013,Alexander:2016,Kubo:2017,Shimakawa:2018}. 

\footnotetext[1]{Various survey bias and restrictions result in vague and inconsistent 
definitions of the {\it protocluster} in any work. This series of papers refers to 
overdense fields on the scale of $\gtrsim10$, $\sim$1--10, and $\lesssim1$ comoving 
Mpc as large-scale structures, protoclusters, and dense cores (groups) for the target, 
respectively.}

The rapid change of star formation rate (SFR) density in the centres of clusters seems 
to follow $(1+z)^{6\sim8}$ from $z\sim2$ to now 
\citep{Kodama:2001,Clements:2014,Smail:2014,Shimakawa:2014,Kato:2016}. 
Such a drastic variation is not only due to the increase in the number of quenched 
galaxies in clusters at lower redshifts 
\citep{Blanton:2009,Muzzin:2012,Wetzel:2012,Burg:2013,Darvish:2016,Paulino:2018} 
but also due to very active star formation in high-$z$ protoclusters 
\citep{Dannerbauer:2014,Umehata:2015,Tadaki:2015,Wang:2016,Oteo:2017}. 
These populations are complicated to reproduce with the classical semi-analytic models 
\citep{Romeo:2015}. Moreover, past studies have reported protoclusters which host 
large numbers of active galactic nuclei (AGNs) (e.g., 
\citealt{Lehmer:2009,Lehmer:2013,Hennawi:2015,Cai:2017b,Krishnan:2017}, but see 
\citealt{Macuga:2018}) including radio-loud sources 
\citep{Pentericci:2002,Rottgering:2003,Venemans:2007,Hatch:2014}. A few studies have 
investigated the energy injection from central AGNs into the ambient gas surrounding 
high-$z$ (proto-) clusters \citep{Nesvadba:2006,Valentino:2016}. There is no good 
understanding of how large an impact AGNs have on the proto-intercluster medium of 
protocluster members. This uncertainty makes it even more difficult to understand the 
mechanism behind the difference in star formation histories in and outside cluster centres. 

It is, therefore, important to characterise massive galaxies in protoclusters. Our 
MAHALO-Subaru (Mapping H-Alpha and Lines of Oxygen with Subaru; \citealt{Kodama:2013}) 
surveys have extensively studied star formation in high-$z$ clusters and 
protoclusters. High-density sampling of line emitters at limited redshift ranges 
($\pm2000$ km~s$^{-1}$) with narrow-band filters have found the inside-out propagation 
of star formation and mapped bottom-up structure growth based on the spatial 
distributions of emission line galaxies and their physical properties. We have 
identified that the regions dominated by bright line emitters are shifted from the 
densest cluster cores to lower-density outskirts and filamentary outer structures, on
timescales from $z\sim3$ to present (e.g., 
\citealt{Hayashi:2010,Koyama:2010,Koyama:2011,Tadaki:2012,Hayashi:2012,Koyama:2013a}). 

Recent deep follow-up \ha\ imaging towards a young protocluster, USS~1558$-$003 at 
$z=2.53$, finds enhanced star formation and concentration of massive \ha\ emitters 
(HAEs) in fragmented group cores \citep{Shimakawa:2018}. 
Furthermore, a follow-up sub-mm/radio campaign with ALMA has shown gas-depleted 
massive galaxies in the very centre of an X-ray cluster, XMMXCS~J2215.9$-$1738 at 
$z=1.46$ \citep{Hayashi:2017}. Their typical gas fraction is no more than 10 per cent 
as opposed to gas-rich sources in the outer regions with gas fractions of $\gtrsim50$ 
percent (\citealt{Hayashi:2018}, see also \citealt{Noble:2017}). Such a sharp contrast 
in time and radial distribution would require e.g., a strong quenching mechanism like 
AGN feedback 
\citep{Springel:2005b,Sijacki:2007,Fabjan:2010,McCarthy:2010,Barnes:2017}, and/or 
rapid gas consumption via starbursts 
\citep{Hopkins:2009,Hayward:2011,Hopkins:2013,Narayanan:2015}. 

Here, in the second part of our MAHALO-Deep cluster survey (MDCS), we investigate the 
properties of massive galaxies in a protocluster associated with a radio galaxy, 
PKS~1138$-$262 at $z=2.16$. This protocluster is known to have an apparent red 
sequence \citep{Kurk:2004a,Kodama:2007,Tanaka:2013}; at the same time, there is a 
strong excess of red \ha-emitting galaxies \citep{Koyama:2013a}. \citet{Koyama:2013a} 
also found that higher fractions of redder and more massive HAEs in higher-density 
regions than in under dense regions in the protocluster, implying that the build-up of 
stellar mass has mostly completed for massive galaxies in the densest parts of the 
protocluster at this time (see also \citealt{Doherty:2010,Hatch:2011,Tanaka:2013}). 
These unique trends suggest that the protocluster is in a critical transition phase 
from young, fragmented, protoclusters, to the classical X-ray clusters at 
$z\lesssim2$. The primary goal of this paper is to determine the stellar mass function 
of protocluster members and then quantify passive fraction and AGN fraction as a 
function of stellar mass. Also, based on multi-wavelength datasets from literature, we 
investigate local number densities, rest-frame colours and SFRs for individual HAEs 
and check if properties of HAEs are different from the field. These will enable us to 
investigate how galaxies in the protocluster stop forming stars. 

We assume the cosmological parameters of $\Omega_M=0.3$, $\Omega_\Lambda=0.7$ and 
$h=0.7$ and adopt a \citet{Chabrier:2003} stellar initial mass function. The AB 
magnitude system \citep{Oke:1983} is employed throughout the Paper.


\section{Target and dataset}\label{s2}


\subsection{PKS~1138$-$262}\label{s2.1}

This paper focuses on a dense protocluster associated with a radio galaxy,  
PKS~1138$-$262 (or MRC~1138$-$262, $\alpha_\mathrm{J2000}=$ 
11$^\mathrm{h}$40$^\mathrm{m}$48$^\mathrm{s}$, $\delta_\mathrm{J2000}=$ 
$-$26$^\mathrm{d}$29$^\mathrm{m}$09$^\mathrm{s}$, 
\citealt{Bolton:1979,Roettgering:1994,Roettgering:1997,Carilli:1997}) at $z=2.156$ 
known as the Spiderweb galaxy \citep{Pentericci:1998,Miley:2006}. The PKS~1138 
protocluster (hereafter PKS~1138) was first explored by \citet{Pentericci:1997}, 
\citet{Kurk:2000}, and \citet{Pentericci:2000}. PKS~1138, together with the SSA22 
protocluster at $z=3.09$ \citep{Steidel:1998,Steidel:2000}, has been extensively 
studied over a long period. The following is a short summary of previous findings on 
PKS~1138 over the past two decades. 

Since the X-ray gas density of nearby galaxy clusters is correlated with the large 
rotation measures \citep{Taylor:1994}, \citet{Carilli:1997} and 
\citet{Pentericci:1997} have suggested that the Spiderweb galaxy resides in a dense 
cluster environment given its observed very high rotation measure of the polarized 
radio emission (6200 rad~m$^{-2}$; see also \citealt{Athreya:1998}). 

\citet{Kurk:2000} and their series of papers 
\citep{Pentericci:2000,Kurk:2003,Kurk:2004a,Kurk:2004b,Croft:2005} identified a 
significant overdensity in this field based on imaging and follow-up spectroscopic 
searches towards \lya\ and \ha\ emitters (LAEs and HAEs, respectively) and distant 
red galaxies (DRGs). Their narrow-band imaging surveys succeeded in selecting 50 LAE 
and 40 HAE candidates, and then spectroscopically confirmed 14 and 9 objects 
respectively. They also found high concentrations of HAEs and DRGs within 0.5 Mpc 
of the Spiderweb galaxy, which are 4--5 times greater than those outside 
the central region. Such massive overdensities have been subsequently confirmed on 
higher dynamic-scales \citep{Koyama:2013a,Shimakawa:2014} and by comparing with other 
radio galaxy environments \citep{Venemans:2007,Mayo:2012,Galametz:2012}. Spectroscopic 
observations tentatively suggested that the protocluster centre of PKS~1138 may have 
halo mass $\sim10^{14}$ \msun\ and virial radius of 0.5 Mpc 
\citep{Pentericci:2000,Kuiper:2011,Shimakawa:2014} assuming that the system is 
collapsed (but see \citealt{Kuiper:2011}). Such a massive overdensity has the 
potential to grow into a massive, Coma-like, galaxy cluster by the present-day 
\citep{Chiang:2013,Lovell:2018}.


\subsection{Data}\label{s2.2}

\begin{figure}
\centering
\includegraphics[width=0.9\columnwidth]{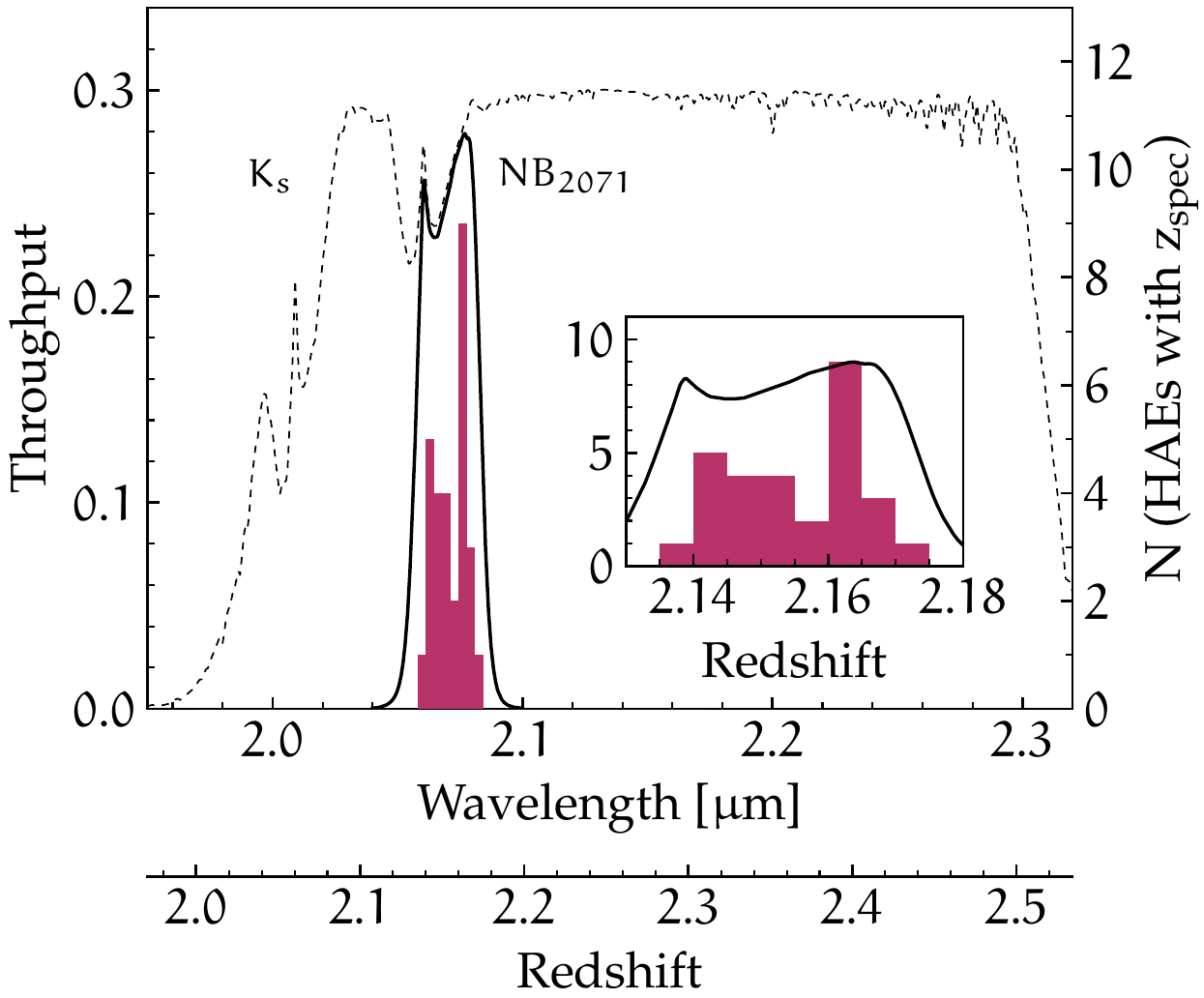}
\caption{System throughputs of $NB_{2071}$ and $K_s$ filters with MOIRCS, represented 
by the black solid and dotted lines, respectively. The red histogram shows spec-$z$ 
distribution of 29 HAEs, 23 of which have been confirmed by \citet{Shimakawa:2014}; 
spectroscopic redshifts for the remainder are taken from the literature 
\citep{Pentericci:2002,Kurk:2004b,Croft:2005,Doherty:2010,Tanaka:2013}. 
One should note that a strong dip at $z\sim2.16$ is caused by the strong OH lines at 
$\lambda=2.0729$ $\mu$m preventing us from spectroscopically identifying the \ha\ line 
of HAEs at this redshift \citep{Shimakawa:2014}.}
\label{fig1}
\end{figure}

We employ the multi broad-band and narrow-band dataset from MDCS and the literature. 
The data consist of $B$, $F475W$, $F814W$, $z'$, $Y$, $J$, $H$, 
$K_{s,\mathrm{MOIRCS}}$ (hereafter $K_s$), $K_{s,\mathrm{HAWKI}}$, and $NB_{2071}$. 
Table~\ref{tab1} summarises the seeing FWHM and limiting magnitudes for these images. 
The $z'$, $J$, $K_s$, and $NB_{2071}$ images are based on the past MAHALO-Subaru 
campaign (S10B-028I, Kodama et al.; \citealt{Koyama:2013a}) and MDCS (S15A-047, Kodama 
et al.). The reduced $B$-band image is provided by Koyama et al. (in preparation), 
and was recently obtained with Suprime-Cam on the Subaru Telescope between May and 
June 2017. The narrow-band filter, $NB_{2071}$ has a central wavelength of 2.071 
$\mu$m with the full width at half-maximum (FWHM) of 270 \AA, which covers the 
\ha-redshift $2.15\pm0.02$ (fig.~\ref{fig1}). 

In addition, we use the reduced Hubble Space Telescope (HST) ACS/WFC data ($F475W$, 
$F814W$), obtained from the Hubble Legacy Archive (HLA), and reduced near-infrared 
(NIR) images ($Y$, $H$, $K_{s,\mathrm{HAWKI}}$) taken with HAWK-I on Very Large 
Telescope (VLT). These original data have been reported in detail by 
\citet{Miley:2006} and \citet{Dannerbauer:2017}, respectively. Moreover, this work 
employs 3.6 and 4.5 $\mu$m IRAC bands \citep{Seymour:2007}. We use the Post-BCD (PBCD) 
products from the Spitzer data archive library. Each IRAC band covers 89 percent of 
the entire narrow-band emitters. The IRAC images are shallow (21.4--21.6 in $3\sigma$ 
limiting magnitude), and we confirmed that has a negligible effect on the measurement 
of physical properties with the SED fitting (\S\ref{s2.4.1}). However, we solely use 
the photometry in these bands to impose restrictions on the rest-frame NIR spectra of 
the targeting HAEs at $z=2.2$; this is crucial when constraining the rest-frame $J$ 
band magnitudes (\S\ref{s3.2}). 

We also introduce here $NB_{2071}$ data, taken as part of the MDCS with the 
Multi-Object Infrared Camera and Spectrograph (MOIRCS; 
\citealt{Ichikawa:2006,Suzuki:2008}) on the Subaru Telescope (the same instrument that 
was used in the past MAHALO-Subaru survey; \citealt{Koyama:2013a}). The observations 
were executed between April 30 and May 6, 2015, under photometric conditions with 
seeing FWHM $\sim0.6$ arcsec. The integration time is 125 min which was split into 180 
sec individual exposures. After combining with the existing $NB_{2071}$ data (186 min 
integration), we reconstructed all the data using the reduction pipeline 
{\sc mcsred}\footnotemark[2] \citep{Tanaka:2011}, which is written as 
{\sc iraf}\footnotemark[3] scripts \citep{Tody:1993}. As described in 
\citet{Shimakawa:2018}, we executed flat fielding, masking objects from the combined 
data in the first run (thus the whole reduction process was conducted twice to remake 
secure object masks), sky subtraction (by median sky and then the polynomially-fitted 
plane for residual sky subtraction), distortion correction, cross-matching, and image 
mosaicing with this pipeline. The reconstructed $NB_{2071}$ image reaches 23.95 mag in 
$3\sigma$ limiting magnitude using a 1.4 arcsec diameter aperture, and its seeing FWHM 
is 0.63 arcsec. The image depth becomes deeper by 0.5 mag than the previous data 
\citep{Koyama:2013a}. The world coordinate system (WCS, 
\citealt{Calabretta:2002,Greisen:2002}) of the narrow-band image is carefully matched 
by the {\sc iraf} scripts ({\sc ccmap} and {\sc ccsetwcs}) to that of the F814W image, 
based on 67 point sources. F814W has one of the best spatial resolutions amongst our 
dataset. The standard deviation of point source separations between the NB$_{2071}$ 
and F814 images suggests that the relative WCS uncertainty would be around 0.04 arcsec 
in the survey area. One should note, however, that the absolute astrometry would have 
0.3 arcsec errors in right ascension and declination based on comparison with the 
Guide Star Catalogue 2 \citep{Lasker:2008}. 

\footnotetext[2]{\url{http://www.naoj.org/staff/ichi/MCSRED/mcsred.html}}
\footnotetext[3]{\url{http://iraf.noao.edu}}

\begin{table}
\centering
\caption{Data summary. The first to fourth columns indicate filter name, 
instrument/telescope, seeing FWHM, $3\sigma$ limiting magnitude in 1.4 arcsec diameter 
aperture including galactic extinction correction, respectively. The fifth column 
shows the galactic extinction based on the NASA Extragalactic Database extinction law 
calculator \citep{Schlegel:1998,Fitzpatrick:1999}$^a$. We employ recalibrated 
estimates from \citet{Schlafly:2011}.
}
\begin{tabular}{lcccc}
\hline
Filter        & Instrument    & FWHM     & 3$\sigma$ & A$\mathrm{_\lambda}$ \\
              & /Telescope    & (arcsec) & (AB)      & (mag)                \\
\hline
$NB_{2071}$   & MOIRCS/Subaru & 0.63     & 23.95     & 0.01  \\
$K_s$         & MOIRCS/Subaru & 0.63     & 23.99     & 0.01  \\
\hline
$B$           & S-Cam/Subaru  & 1.15     & 26.56$^c$ & 0.14  \\
$F814W$       & ACS/HST       & 0.11     & 26.33$^b$ & 0.06  \\
$F475W$       & ACS/HST       & 0.11     & 27.02$^b$ & 0.13  \\
$z'$          & S-Cam/Subaru  & 0.70     & 26.35     & 0.05  \\
$Y$           & HAWK-I/VLT    & 0.37     & 26.08$^b$ & 0.04  \\
$J$           & MOIRCS/Subaru & 0.69     & 24.33     & 0.03  \\
$H$           & HAWK-I/VLT    & 0.49     & 25.11$^b$ & 0.02  \\
$K_s$         & HAWK-I/VLT    & 0.38     & 24.75$^b$ & 0.01  \\
3.6 $\mu$m    & IRAC/Spitzer  & 1.8      & 21.42$^d$ & 0.00  \\
4.5 $\mu$m    & IRAC/Spitzer  & 1.8      & 21.57$^d$ & 0.00  \\
\hline
\multicolumn{5}{l}{$^a$ \url{http://irsa.ipac.caltech.edu/applications/DUST/}}\\
\multicolumn{5}{l}{$^b$ limiting magnitudes after PSF matching with $NB_{2071}$}\\
\multicolumn{5}{l}{$^c$ limiting magnitude in 2.5 arcsec aperture diameter}\\
\multicolumn{5}{l}{$^d$ limiting magnitude in 8.0 arcsec aperture diameter}\\
\end{tabular}
\label{tab1}
\end{table}


\subsection{Sample selection}\label{s2.3}


\subsubsection{Narrow-band selection}\label{s2.3.1}

We selected the sample of HAEs by the combined technique of narrow-band selection 
\citep{Bunker:1995} and $Bz'K_s$ colour selection \citep{Daddi:2004}. 
The former selection is defined by the following criteria, 
\begin{eqnarray}
K_s-NB &>& -2.5\log(1-\Sigma\delta10^{-0.4(ZP-NB)}) +\zeta \label{eq1} \\
K_s-NB &>& 0.253 \label{eq2} 
\end{eqnarray}
where $\Sigma$ is the confidence level (in sigma) of the colour-excess and $\delta$ is 
defined by the combined $1\sigma$ background noise at $NB$ ($\equiv NB_{2071}$) and 
$K_s$ bands, ($\delta=\sqrt{\sigma_{NB}(S)^2+\sigma_{Ks}(S)^2}$ where $S$ is the 
photometric aperture area). $ZP$ is the zero point magnitude of the $NB_{2071}$ image. 
$\zeta$ is a correction factor of the colour term. We use $\zeta=-0.04$ which 
corresponds to the median value of the colour terms in the entire HAEs 
(Appendix~\ref{a1}). The former equation reflects the narrow-band flux limit 
($>3\times10^{-17}$ erg~s$^{-1}$cm$^{-2}$) and the latter colour threshold 
(eq.~\ref{eq2}) corresponds to the equivalent width limit of narrow-band flux 
(EW$_\mathrm{NB}=30$ \AA\ in the rest frame for $z=2.15$). The EW$_\mathrm{NB}$ limit 
is chosen so as not to accidentally pick up contaminant non-emitters 
(Appendix~\ref{a1}). 

We note that the measurement of background noise ($\sigma(S)$) in this work and recent 
other narrow-band studies (e.g., \citealt{Hayashi:2010,Matthee:2017,Hayashi:2018b}) is 
different from the original calculation by \citet{Bunker:1995}. \citet{Bunker:1995} 
define the noise by $\sigma(S)=\sqrt{\pi r^2\sigma_0^2}$ where $r$ is an aperture 
radius and $\sigma_0$ is $1\sigma$ background noise in pixel. This definition assumes 
that photometric error is proportional to the aperture radius, however, the real 
science images have pixel-to-pixel correlations (see e.g., \citealt{Skelton:2014}), 
which lead to underestimated background noise especially in the larger aperture area. 
We indeed obtained the power law functions $N=$ 1.412 and 1.345 ($N$ is defined by 
$\sigma(S)\propto r^N$) at $NB_{2071}$ and $K_s$ images based on randomly-positioned 
empty apertures with different radii across the image. This work thus employs the 
fixed background noise at each band ($NB_{2071,1\sigma}=25.14$ mag and 
$K_{s,1\sigma}=25.18$ mag), derived by placing random empty apertures with the same 
diameter (1.4 arcsec) in the selection process. We here ignore the local sky variance 
that is estimated to be $\le0.1$ mag across each image. 

We performed source detection in the reduced narrow-band image, using SExtractor (ver. 
2.19.5, \citealt{Bertin:1996}). We set detection parameters of {\sc detect$\_$minarea} 
$=9$, {\sc detect$\_$thresh} $=1.2$, {\sc analysis$\_$thresh} $=1.2$, and 
{\sc deblen$\_$mincont} $=1\times10^{-4}$. $K_s$-band photometry was conducted by the 
double-image mode of the SExtractor with the $NB_{2071}$ image for the source 
detection. Input parameters for source photometry are set with {\sc back$\_$size} 
$=64$, {\sc back$\_$filtersize} $=5$, {\sc backphoto$\_$type} = {\sc local}, and 
{\sc backphoto$\_$thick} $=32$ (the same applies all source photometric processes 
hereafter). According to the Monte Carlo simulation with randomly-positioned PSF models 
embedded in the narrow-band image (see Appendix~\ref{a2} for details), the magnitude 
limit of 95 percent completeness in the source detection is $\sim22.8$ mag. 

\begin{figure}
\centering
\includegraphics[width=0.9\columnwidth]{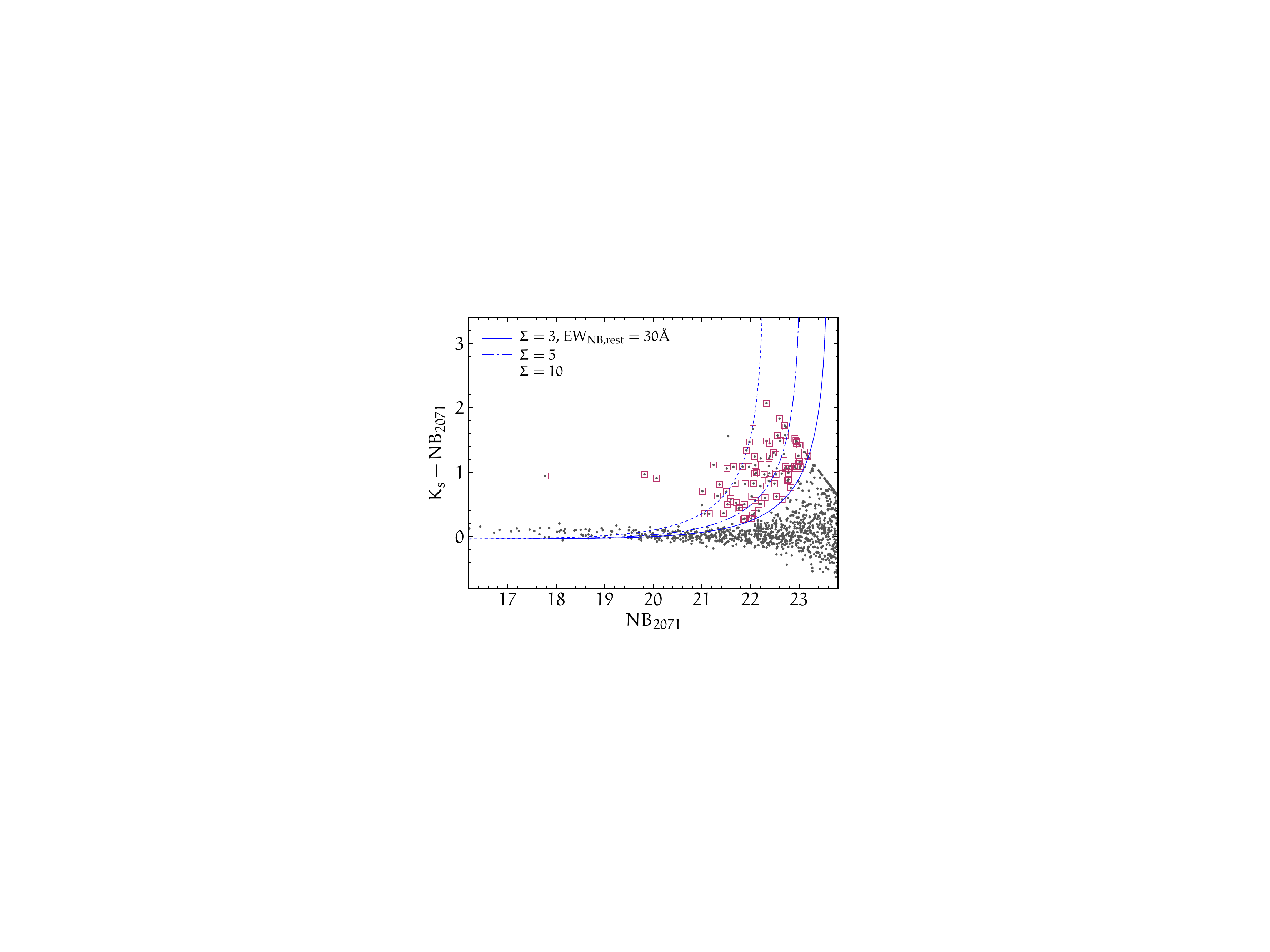}
\caption{Colour--magnitude diagram, $NB_{2071}$ versus $K_s-NB_{2071}$.
The black dots are all NB-detected sources. The red squares indicate narrow-band 
emitters showing their narrow-band flux excesses greater than three sigma levels and 
EW$_\mathrm{NB}$ higher than 94.5 \AA\ (30 \AA\ in the rest frame at $z=2.15$). 
Blue solid, dash-dotted, and dotted lines are 3, 5, and 10 $\Sigma$ excess, 
respectively. The blue horizontal line shows the EW limit.}
\label{fig2}
\end{figure}

Figure~\ref{fig2} shows the colour--magnitude diagram ($NB_{2071}$ versus 
$K_s-NB_{2071}$) for $NB_{2071}$ detected sources in the PKS~1138 region. One should 
note here that their $NB_{2071}$ magnitudes and colours are based on the fixed 
aperture photometry of 1.4 arcsec diameter. We assume two sigma limiting magnitude for 
non-detections at $K_s$ band. We then select the objects with $\Sigma>3$ 
colour-excesses as our narrow-band emitter (NBE) sample. The $\Sigma=3$ limit in this 
work is more conservative when compared to $\Sigma=2$ in \citet{Kurk:2004a} and 
$\Sigma=2.5$ in \citet{Koyama:2013a} for the same field. Nevertheless, thanks to the 
deeper observing depth than the previous work, $\gtrsim1.5$ times more emitters (97 
samples) meet our colour criteria in the same area. Our Monte Carlo simulation claims 
that this narrow-band selection has 68 and 95 percent completeness at 
$NB_{2071}=$ 22.80 and 22.45 mag, respectively (Appendix~\ref{a2}). 

\subsubsection{Colour selection}\label{s2.3.2}

Combined with the past spectroscopic observations 
\citep{Pentericci:2002,Kurk:2004b,Croft:2005,Doherty:2010,Tanaka:2013,Shimakawa:2014} 
and narrow-band \lya\ imaging \citep{Kurk:2000}, we already have 36 secure HAE sources 
with spectroscopic confirmation or narrow-band excess in two filters in \ha\ and \lya\ 
lines. For the remaining NBEs, even though the survey field is known to be a massive 
overdense region, it is important to carry out further selection to select HAEs more 
likely to be associated with PKS~1138 at $z=2.2$, and exclude other line contaminants 
e.g., background \oiii, \hb\ line emitters at $z>3$ and foreground \paa\ emitter at 
$z=0.1$. 

\begin{figure}
\centering
\includegraphics[width=0.9\columnwidth]{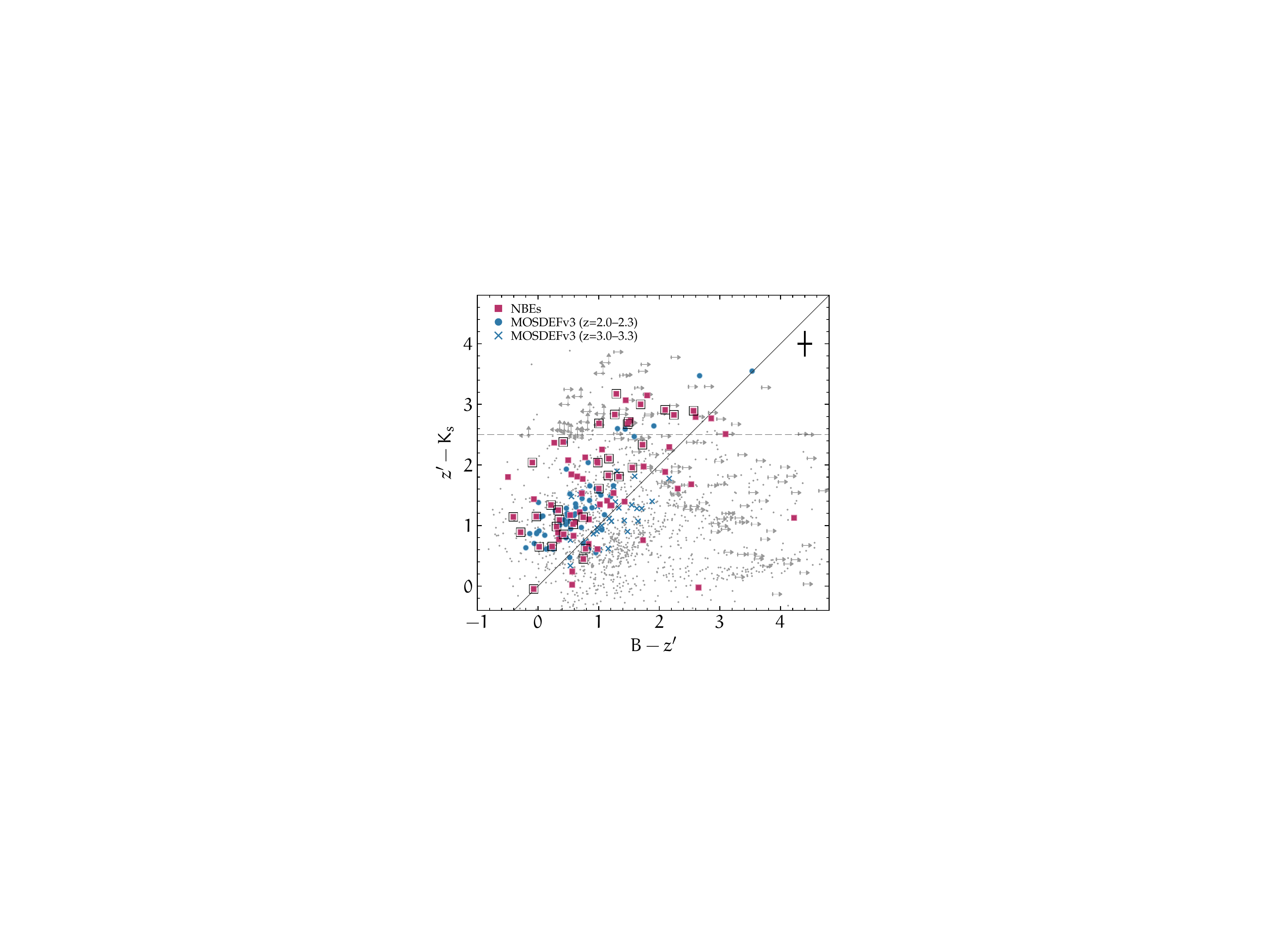}
\caption{$Bz'K_s$ colour--colour diagram for the PKS~1138 region. Red squares and blue 
symbols indicate the narrow-band emitters and the spec-$z$ samples at $z=$ 2.0--2.3 
(circles) and 3.0--3.3 (crosses) from the MOSDEF survey \citep{Kriek:2015}, 
respectively. Spectroscopically-confirmed HAEs are highlighted by open black squares. 
Grey dots are NB-detected sources. The figure only shows objects with $>2\sigma$ 
detection at $K_s$-band. Two sigma limiting magnitudes replace band photometry for 
faint sources at $B$ or $z'$-band. The black solid line is our colour threshold defined 
to remove those foreground or background contaminants. The horizontal dashed line is 
the colour criterion of DRG. The black cross on the upper right shows the typical 
$1\sigma$ photometric error of the narrow-band emitters.} 
\label{fig3}
\end{figure}

Colour--colour selection has been widely used for further selection to remove other 
line contaminants \citep{Koyama:2013a,Tadaki:2013}. Although it would be better to also
integrate with photometric redshifts as demonstrated by the High-redshift(Z) Emission 
Line Survey (HiZELS; \citealt{Geach:2008,Sobral:2013}), this work does not employ 
photometric redshifts since available photometric bands are not many as used in 
such large panoramic surveys. 
For $z\sim2.2$ sources, we employ the well known $BzK$ ($\equiv(z-K_s)-(B-z)$) 
selection \citep{Daddi:2004,Daddi:2005} which is accessible given our imaging dataset 
($Bz'K_s$) as employed in the previous work \citep{Koyama:2013a}. The $BzK$ colour 
criteria enable culling of star-forming and passive galaxies at $z\sim$ 1.4--2.5 
without extinction correction. We plot NB detections with $>2\sigma$ detection at $K_s$ 
band on the $Bz'K_s$ colour--colour diagram (fig.~\ref{fig3}). In addition, we also 
show spec-$z$ sources in the COSMOS--CANDELS field 
\citep{Scoville:2007,Capak:2007,Grogin:2011,Koekemoer:2011} from the MOSFIRE Deep 
Evolution Field survey (MOSDEF; \citealt{Kriek:2015}) as a reference sample. $Bz'K_s$ 
colours of these spec-$z$ sources are derived from the 3D-HST database 
\citep{Brammer:2012,Skelton:2014}, which were originally taken by the large legacy 
surveys with the Subaru Telescope and Vista \citep{Taniguchi:2007,McCracken:2012}.

We derive $Bz'K_s$ colours of NB-detections based on the output from the SExtractor 
({\sc mag$\_$auto}) with double image mode. All images were tailored to the size of 
the narrow-band image with a scale of 0.117 arcsec per pixel. We employ the outputs of 
{\sc mag$\_$auto} in each band and we set {\sc kron$\_$fact} $=2.5$. One should note 
that source photometry for the $B$-band image with a much larger seeing FWHM 
(table~\ref{tab1}) was also executed independently and then we chose brighter $B$-band 
flux densities from single or double image mode for individual sources. Based on 
colours of confirmed members and spec-$z$ sources, we set the colour thresholds of 
$Bz'K_s>0$ or ($z'-K_s$) $>2.5$, and then select an additional 32 HAE members as well 
as remove 16 narrow-band emitters as other line emitters (table~\ref{tab2}). Here, we 
adopt a $Bz'K_s$ colour criterion that is different from the \citet{Daddi:2004} 
prescription ($BzK_s\geq-0.2$). We assume two sigma limiting magnitudes for 
non-detections, and then evaluate them if those upper limits or lower limits can meet 
our selection criteria. The $Bz'K_s$ selection cannot perfectly guarantee that the 
selected NBEs are our targeting HAEs at $z=2.15\pm0.02$ though (fig.~\ref{fig3}). We 
indeed find that two confirmed protocluster members drop below our selection limit. On 
the other hand, some reference spec-$z$ sources at $z=$ 3.0--3.3 break into the realm 
of $Bz'K_s$-selected galaxies. Also, colour-selected HAEs and rejected line emitters 
near boundaries of the colour criteria cannot be securely classified once we take 
account of those photometric errors. While we count these colour-selected emitters as 
HAEs throughout this paper at this time, we definitely require follow-up spectroscopy 
for the robust identification of these sources in the future. The online catalogue 
(appendix~\ref{a3}) summarises the identification status for individual HAE samples in 
detail. 

When taken together, a total number of 68 HAEs have been selected as the protocluster 
members in this work. 51 out of them are already discovered by our previous survey 
\citep{Koyama:2013a}, meaning that our deeper data increase the number of the HAE 
sample by 33 percent. Besides these, we have 13 NBEs which cannot be removed by the 
$Bz'K_s$ colour due to insufficient photometric data. These unknown emitter samples 
are defined as HAE candidates. Given the ten times higher density in the protocluster 
region (\S\ref{s3.4}), most of these faint emitters should be HAEs. The contamination 
rate in the HAE candidates may be around $\sim20$ percent considering 16 
colour-rejected NBEs amongst the 84 ($36+32+16$) emitters (table~\ref{tab2}).

\begin{table}
\caption{Classification of NBEs and DRG$_\mathrm{nIR}$ samples in PKS~1138 region. This 
work employs confirmed and colour-selected emitters as the HAE sample. We also use HAE 
candidates when we derive distribution functions (\S\ref{s3.4}). See \S\ref{s2.3} for 
details.}
\begin{center}
\begin{tabular}{lrl}
\hline
Class                  & N   & Description \\
\hline
NBEs                   &  97 & narrow-band emitters ($\Sigma>3$) \\
HAEs (confirmed)       &  36 & confirmed by spec-$z$ or \lya\ line \\
HAEs (by colours)      &  32 & selected by $Bz'K_s$ colour \\
other line emitters    &  16 & rejected by $Bz'K_s$ colour \\
\hline
HAEs                   &  68 & confirmed $+$ colour-selected HAEs \\
HAE candidates         &  13 & cannot be rejected by colours \\
\hline
DRG$_\mathrm{nIR}$     &  34 & $z'-K_s>2.5$ w/o 24 $\mu$m detection \\
\hline
\end{tabular}
\end{center}
\label{tab2}
\end{table}%


\subsubsection{Distant red galaxies (DRGs)}\label{s2.3.3}

We also establish a reference sample of distant red galaxies (DRGs) that do not show 
signs of active star formation. These objects allow us to infer the selection bias of 
our narrow-band technique at the massive end, and also provide the upper limit to the 
quiescent population in the derivation of the stellar mass function (\S\ref{s3.1}). We 
first chose objects with significant $K_s$-band detection, $K_s<23.4$ ($5\sigma$ limit 
mag), corresponding to the 95 percent completeness limit for massive galaxies 
(M$_\star>10^{10.5}$ \msun) according to the photo-$z$ source catalogue in the COSMOS 
field \citep{Laigle:2016}. We then select passive $BzK$ (pBzK) galaxies that satisfy 
($z'-K_s$) $>2.5$ (fig.~\ref{fig3}) and do not overlap with NBEs nor MIPS/Spitzer 24 
$\mu$m sources reported by \citet{Koyama:2013a}. One should note that this colour 
threshold is different from the classic definition of DRGs ($J-K_s>2.3$ in vega) by 
\citet{Dokkum:2004} and \citet{Franx:2003}. The cross-checking with MIPS 24$\mu$m 
sources allows us to remove significant dusty starburst populations. The detection 
limit at the MIPS 24$\mu$m image roughly corresponds to the infrared luminosity of 
$L_\mathrm{IR}\sim10^{12}$ $L_\odot$ and SFR $\sim100$ \msun~yr$^{-1}$ at $z=2.15$.

This selection results in 34 DRG$_\mathrm{nIR}$ candidates without bright-IR emission 
($L_\mathrm{IR}\gtrsim10^{12}$ $L_\odot$), which are described as DRG$_\mathrm{nIR}$ 
hereafter. Three of these are known to be protocluster members confirmed with 
spectrophotometric analysis \citep{Tanaka:2013}. According to a photometric redshift 
code, {\sc eazy} \citep{Brammer:2008,Brammer:2011}, measured photometric redshifts 
fall within $z=2.1\pm0.2$ in 17 sources.


\subsubsection{X-ray sources}\label{s2.3.4}

We checked the presence of X-ray emission from our HAE samples using an image from the 
Chandra X-ray Observatory. Our survey field is covered by the S3 chip with the ACIS-S 
detector. The data quality and source catalogue were published in \citet{Carilli:2002} 
and \citet{Pentericci:2002}. However, we double-checked the data independently based 
on the Chandra Source Catalogue (CSC v1.1, \citealt{Evans:2010}) and also by analysing 
the original data with the Chandra Interactive Analysis of Observations (CIAO v4.7.6) 
to obtain more detailed coordinates. 

Based on X-ray detections selected by the CIAO code {\sc wavdetect} for an 
exposure-weighted reduced image with {\sc mkexpmap}, we found that six HAEs 
(\#40,46,58,68,73,95) have X-ray detections within 0.4 arcsec separation angle at 
higher than four sigma levels. The faintest X-ray source has $4\times10^{-15}$ 
erg~s$^{-1}$cm$^{-2}$ and $1.4\times10^{44}$ erg~s$^{-1}$ in unabsorbed flux and 
luminosity (assuming the redshift of $z=2.15$) at the broadband (0.5--7.0 keV) 
according to the CSC, respectively. Given such a shallow detection limit, these X-ray 
sources are expected to originate from active galactic nuclei. All of these X-ray 
sources have been identified as \#3,5,6,7,16 in \citet{Pentericci:2002}, whereas \#7 
contains two HAE sources defined in this work: one is the Spiderweb radio galaxy (\#73 
in this work), and the other is HAE-058. Corresponding identification numbers to each 
HAE are fully described in our catalogue (appendix~\ref{a3}).


\subsubsection{Other resources}\label{s2.3.5}

The Spiderweb protocluster is a well-surveyed region, with numerous studies in 
addition to those already mentioned, e.g., MIPS 24 $\mu$m imaging with 
the Spitzer Space Telescope \citep{Mayo:2012,Koyama:2013a,Koyama:2013b}, LABOCA 870 
$\mu$m imaging with the APEX telescope \citep{Dannerbauer:2014}, CO($1-0$) observation 
with ATCA \citep{Emonts:2016,Dannerbauer:2017,Emonts:2018}, and CO($3-2$) observation 
with ALMA (Tadaki et al. in preparation). 

Because of the restricted field coverage relative to our survey area, or serious 
blending issue due to poor spatial resolutions, we do not use these other resources, 
mostly in the mid-IR to radio regime, unless otherwise mentioned. On the other hand, 
these past studies are useful to characterise some specific HAEs, and thus, such 
information is referenced where appropriate throughout the paper.


\subsection{Derivation of physical properties}\label{s2.4}

This section explains how we derive line flux, stellar mass, and amount of dust 
reddening. The measuring methods are similar to those in the first paper of the 
MDCS series \citep{Shimakawa:2018}.


\subsubsection{SED fitting}\label{s2.4.1}

We use SED-fitting to derive stellar masses and dust extinctions of our samples based 
on the SED-fitting code ({\sc fast}) distributed by \citet{Kriek:2009}. We use the 
\citet{Bruzual:2003} stellar population model, the \citet{Calzetti:2000} extinction 
law, and the \citet{Chabrier:2003} IMF. We then run the code with a fixed redshift of 
$z=2.15$ independently of spectroscopic confirmation and low metal abundance of 
$Z=0.004$ ($0.2 Z_\odot$), and assume delayed exponentially declining star formation 
history (SFR $\propto~t\cdot exp(-t/\tau)$). $\tau$ value and age are allowed to be 
$10^9$--$10^{11}$ yr and $10^{7.6}$--$10^{9.4}$ yr, respectively. We allow the amount 
of stellar extinction ($A_V$) to be between 0 and 3 mag. The outcome of the choice of 
these parameter sets does not significantly affect the stellar mass estimations. 
However, the obtained dust extinction systematically depends on input parameters. If 
we employ solar metal abundance instead of $Z=0.2 Z_\odot$, for instance, derived 
$A_V$ values become systematically lower by 0.2--0.3 mag than those with $Z=0.004$. 
Indeed, dust correction is the major issue for narrow-band studies due to the lack of 
\hb\ line information. Considering this model dependency, we pay special attention to 
physical properties requiring dust correction, such as SFR, throughout the paper. 

We then carried out the SED fitting based on multi-band photometry derived in the same 
way as for the $Bz'K_s$ colour estimation in the previous subsection. First, we 
performed PSF-matching for the $F475W$, $F814W$, $Y$, $H$, $K_{s,\mathrm{HAWKI}}$ 
images to the seeing size of $NB_{2071}$. Source photometry at IRAC 3.6 and 4.5 $\mu$m 
bands were conducted independently, and we cross-matched those to NBEs within 1 arcsec 
distance. Whilst we executed the SED-fitting with IRAC photometry if available, we 
confirm that IRAC data have negligible effects on stellar mass and $A_v$ estimation in 
our samples. This lack of systematic discrepancy regardless of the availability of 
IRAC photometry is consistent with past work \citep{Elsner:2008,Muzzin:2009}.  
Derived stellar masses and dust reddening are summarised in appendix~\ref{a3}. We 
employ $1\sigma$ errors of obtained parameters from 100 Monte Carlo simulations 
carried out with the {\sc fast} code. 

\begin{figure}
\centering
\includegraphics[width=0.9\columnwidth]{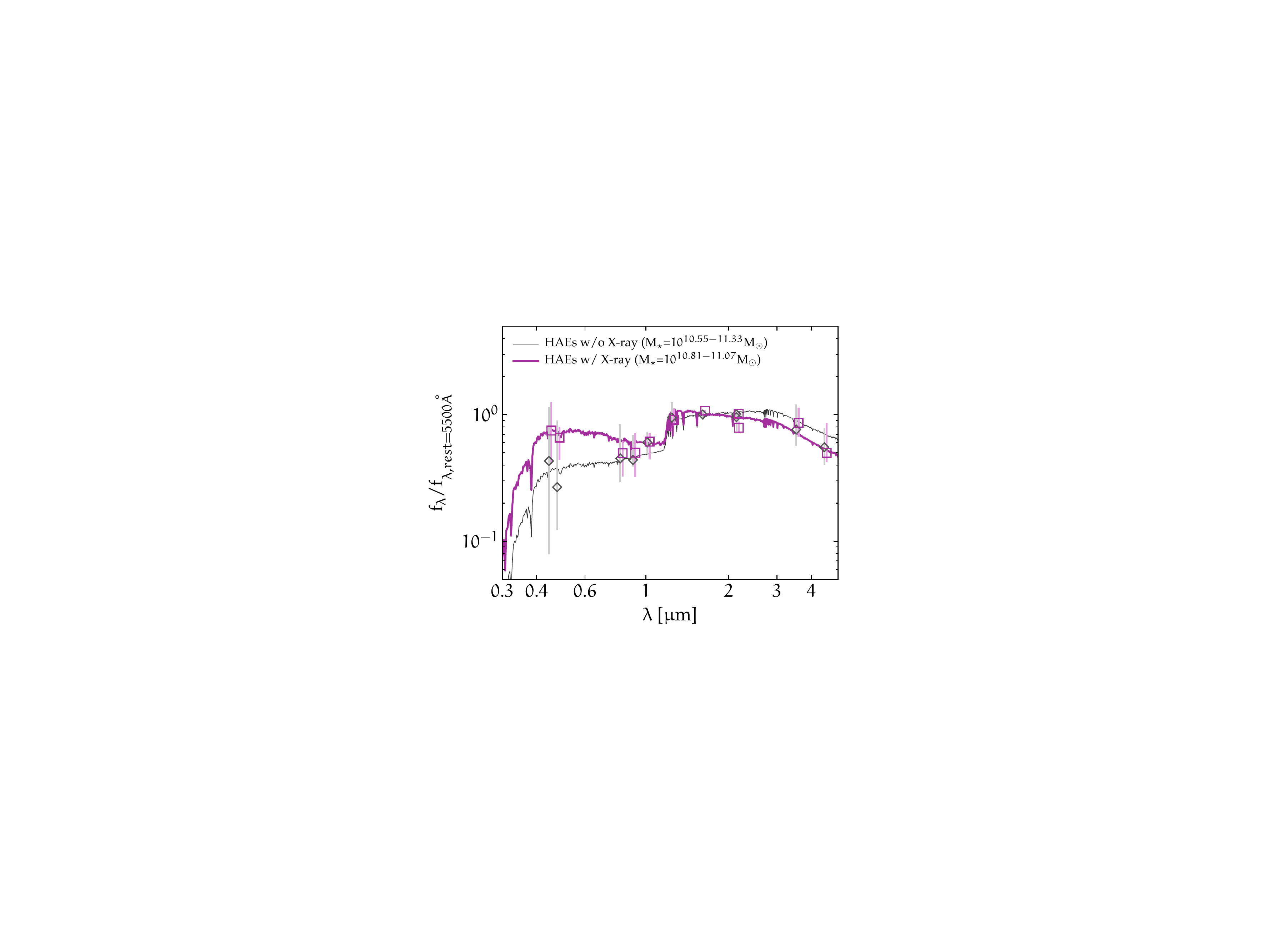}
\caption{
Median stellar spectra of massive (M$_\star>10^{10.5}$ M$_\odot$) HAEs with 
and without X-ray emission, which are represented by purple thick and black thin 
lines, respectively. These are derived from the median values of fitted SED spectra 
normalised at $\lambda_\mathrm{rest}=5500$ \AA\ for individuals. Their median values 
and $1\sigma$ scatters of observed flux densities at 11 photometric bands are shown by 
purple squares and black diamonds, respectively (the data points are slightly shifted 
in a transverse direction for better visibility). Minimum--median--maximum values of 
derived log stellar mass in each group are 10.81--11.01--11.07 in the HAEs with X-ray 
sources and 10.55--10.77--11.33 in the HAEs without X-ray counterparts.}
\label{fig4}
\end{figure}

Typical fitted SED spectra of massive HAEs (M$_\star>10^{10.5}$ M$_\odot$) are 
presented in fig.~\ref{fig4}, which are divided into the spectra of HAEs with or 
without X-ray counterparts. We should note that, in spite of the importance of the SED 
decomposition into the stellar light and the nuclear component 
\citep{Merloni:2010,Santini:2012}, this work ignores this procedure due to the lack of 
photometric bands at rest-frame IR bands. Although we have Spitzer/MIPS 24 $\mu$m 
data, the serious blending issue does not provide us with reliable mid-IR photometry. 
At the least, the stellar mass measurement of the Spiderweb radio galaxy, the most 
luminous X-ray source in our sample, is highly uncertain. Also, we see a clear excess 
at IRAC bands from the extrapolation of model-inferred SED in one of X-ray HAEs (\#58) 
and whose derived stellar mass should be overestimated as well. For other AGN host 
HAEs, since their photometry can be fitted only by stellar components, it remains 
unclear how reliable our stellar mass estimates are. High resolution deep mid-IR data, 
by e.g., JWST/MIRI \citep{Rieke:2015}, are needed to decompose those SEDs and obtain 
pure stellar components.

\subsubsection{Narrow-band flux}\label{s2.4.2}

We obtained narrow-band line flux ($F_\mathrm{NB}$), emission subtracted flux 
density at $K_s$-band ($f_c$), and rest-frame equivalent width of narrow-band flux 
(EW$_\mathrm{NB}$) by the following formula, 
\begin{eqnarray}
F_\mathrm{NB} &=& \Delta_\mathrm{NB} \frac{f_\mathrm{NB}-f_\mathrm{Ks'}}{1-\Delta_\mathrm{NB}/\Delta_\mathrm{Ks}} \label{eq3} \\
f_{c} &=& \frac{f_\mathrm{Ks'}-f_\mathrm{NB}\cdot\Delta_\mathrm{NB}/\Delta_\mathrm{Ks}}{1-\Delta_\mathrm{NB}/\Delta_\mathrm{Ks}} \label{eq4} \\
\mathrm{EW_{NB}} &=& \frac{F_\mathrm{NB}}{f_c\cdot(1-z)} \label{eq5}
\end{eqnarray}
where $\Delta_\mathrm{NB}$ and $\Delta_\mathrm{Ks}$ are full-width-half-maximum (FWHM) 
of NB$_{2071}$ (270 \AA) and $K_s$ band (3100 \AA) filters, respectively. 
$f_\mathrm{Ks'}$ is $K_s$-band flux density including the colour term correction 
(\S\ref{s2.3.1}). We employed $-0.04$ mag for the colour term correction based on the 
median value of model-inferred SED spectra of the entire HAEs (fig.~\ref{fig1a}). Given 
the similar centre wavelength between the two filters, uncertainty from the colour 
correction is negligibly small relative to the total flux errors. 

On the other hand, the shape of the filter throughput including atmospheric 
transmission on Maunakea (fig.~\ref{fig1}) may cause an additional $\sim12$ percent 
error in the flux estimation according to the standard deviation of the response curve 
at wavelengths within the filter FWHM. We incorporate this error budget into the 
narrow-band flux errors of HAEs individually. One should note that we likely 
overestimate this error value given that HAEs tend to gather towards the protocluster 
system ($z\sim2.156$) along the line of sight \citep{Shimakawa:2014}. 

We then obtained observed \ha\ luminosities of HAEs as follows. We assume a fixed 
redshift of $z=2.15$ that corresponds to the \ha\ redshift captured by the centre of 
the $NB_{2071}$ filter. Flux contribution from the \nii$\lambda\lambda6550,6585$ 
doublet is corrected based on their stellar masses derived by the SED fitting 
(\S\ref{s2.4.1}). Our past spectroscopic observation has derived typical 
\nii$\lambda6585$/\ha\ flux ratios (N2; \citealt{Pettini:2004}) for HAEs in PKS~1138 
at different stellar mass bins \citep{Shimakawa:2015}, which can be characterised by 
the following relation, 
\begin{equation}
\mathrm{N2} = -0.71 +0.33 \times [\log(\mathrm{M_\star/M_\odot}) - 10]. 
\label{eq6}
\end{equation}
This prescription enables relatively self-consistent N2 correction for the narrow-band 
flux in our sample. We assume \nii$\lambda6550$:\nii$\lambda6585$ = 1:3 to remove 
\nii$\lambda6550$ flux as well \citep{Osterbrock:1974}. In addition, we incorporate 
the uncertainty of the N2 correction into the derived \ha\ luminosities based on the 
typical observational scatter $\Delta\log\mathrm{(O/H)_{N2}}\sim0.1$ dex (i.e., 
$\Delta$N2 $=0.18$ dex) of the N2-inferred mass--metallicity relation 
\citep{Tremonti:2004,Mannucci:2010,Steidel:2014}. The calculated \ha\ luminosities and 
those total error budgets are summarised in the online catalogue (appendix~\ref{a3}).


\section{Results}\label{s3}

The goal of this paper is to investigate physical properties of massive HAEs in the 
Spiderweb protocluster (PKS~1138) at $z=2.2$. Our previous paper 
\citep{Shimakawa:2018} reported the vigorous formation of more massive galaxies in 
fragmented dense groups alongside intergroup regions within the USS~1558 protocluster 
at $z=2.53$. Compared to USS~1558, PKS~1138 is apparently a more advanced and reddened 
protocluster system  \citep{Kodama:2007}. More specifically, \citet{Galametz:2012} 
have reported three times higher number density of old populations selected by IRAC 
colour in this field compared to the typical radio galaxy environments at high 
redshifts including USS~1558. Given the fact that massive galaxies in the protocluster 
are destined to grow into bright red sequence galaxies in the local Universe, 
identifying these massive HAEs will help us to infer the evolutionary steps cluster 
galaxies would have experienced in their maturing phases at $z\sim2$.


\subsection{Stellar mass functions}\label{s3.1}

We first derive the stellar mass function of HAEs in PKS~1138 whose stellar masses are 
individually derived from the SED fitting. 
Since the mass estimations do not include the SED decomposition to remove AGN 
contamination, all results obtained in this section must be taken with caution. 
Analysing the Spiderweb protocluster region especially suffers from this issue due to 
the number excess of luminous X-ray sources \citep{Pentericci:2002}. 

We use the same measuring method as in our previous paper \citep{Shimakawa:2018} for 
HAEs in USS~1558 protocluster region at $z=2.53$. 
The most important part of the derivation of the stellar mass function for narrow-band 
selected emitters is the completeness correction. Following \citet{Shimakawa:2018}, we 
evaluate both detection completeness and selection completeness with the Monte Carlo 
simulation. The detection completeness is defined as the fraction of missing samples 
in the source detection process, which is highly dependent on the initial parameters 
of the SExtractor code (\S\ref{s2.3.1}). The selection completeness is a specific 
problem of the narrow-band selection, which is firstly noted by \citet{Sobral:2009} 
and then developed by their following analyses 
\citep{Sobral:2012,Sobral:2013,Sobral:2014}. The selection completeness indicates the 
completeness in the process of the narrow-band colour selection (\S\ref{s2.3.1}). 
Evaluating the selection completeness is especially crucial since the narrow-band 
selection is not only based on the depth of the narrow-band image, but is also 
dependent on the colour between narrow-band and broad-band photometry (see 
\citealt{Sobral:2009}). Indeed, our Monte Carlo analysis indicates that the 
narrow-band selection requires an additional 20--70 percent completeness correction at 
the faint end relative to the completeness correction only for the detection. The 
detailed procedure of our completeness correction is examined in Appendix~\ref{a2}.  

Figure~\ref{fig5} shows the number densities as a function of stellar mass for HAEs in 
PKS~1138. We evaluate the number densities of HAEs by 
$\phi(\log{L})=\Sigma_i(V_\mathrm{max}\cdot C(NB)\cdot\Delta(\log{L}))^{-1}$, where 
$L=\mathrm{M}_\star/\mathrm{M}_\odot$ and $V_\mathrm{max}$ ($=3676$ co-Mpc$^3$) is the 
volume size, respectively. The latter is obtained from the filter FWHM of the 
narrow-band filter (55 co-Mpc) and the survey area (66 co-Mpc$^2$). Since the redshift 
distribution of HAEs is more concentrated around the protocluster centre 
\citep{Shimakawa:2014}, we tend to overestimate the volume size. Open squares in 
fig.~\ref{fig5} include the completeness corrections that also incorporate 13 
unclassified HAE candidates (table~\ref{tab2}) to compensate loss from the $Bz'K_s$ 
colour selection. We count these candidates with the additional correction of 20 
percent possible contamination (\S\ref{s2.3.2}). One should note that while this 
further correction could affect the resultant fitting parameters, this does not change 
our conclusion since all HAE candidates are less-massive galaxies at stellar masses 
lower than $10^{10.5}$ \msun, most of which are outside of the scope of this work.

\begin{figure}
\centering
\includegraphics[width=0.95\columnwidth]{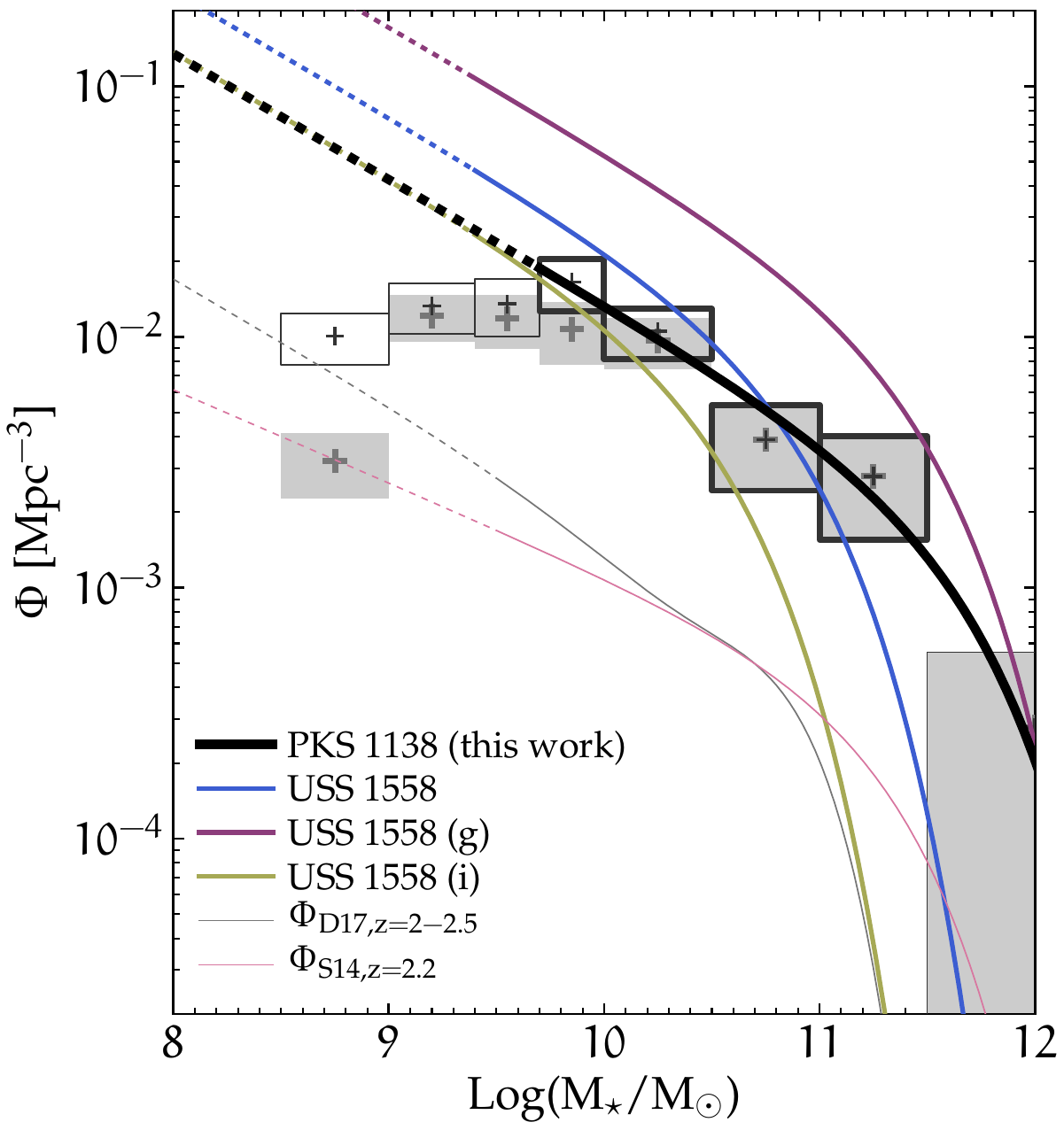}
\caption{Stellar mass functions in various fields at $z\sim2$. Open and filled boxes 
show the number densities with and without completeness correction at each stellar 
mass bin. The blue curve represents the stellar mass function in the USS~1558 
protocluster at $z=2.5$. Purple and yellow curves are those in the group and 
intergroup regions therein, respectively. Thin grey and pink lines are the stellar 
mass function in the general field at $z\sim2$. The former is based on star-forming 
galaxies selected from their rest-$UVJ$ colours and photo-$z$ \citep{Davidzon:2017}. 
The latter is narrow-band selected HAEs including additional sorting with photo-$z$ 
and colours \citep{Sobral:2014}. Dotted lines indicate the extrapolated lines from 
the available data range in each function. 
}
\label{fig5}
\end{figure}

We fit the stellar mass distribution using the Schechter function 
\citep{Schechter:1976}, which is given by the following equations, 
\begin{eqnarray}
\phi(L)dL &=& \phi^\ast (\frac{L}{L^\ast})^\alpha \exp(-\frac{L}{L^\ast})\frac{dL}{L^\ast} ~~~~~~~~~~~~~~~~ or \\
\phi(L)dL &=& \phi^\ast (\frac{L}{L^\ast})^{\alpha+1} \exp(-\frac{L}{L^\ast})\ln{10}~d(\log{L}),
\end{eqnarray}
where $L^\ast$ is the characteristic stellar mass at which the power law slope cuts 
off. We then fit the stellar mass distribution with the Schecter function based on the 
{\sc mpfit} code \citep{Markwardt:2009}\footnotemark[4]. We do not use the stellar 
mass bins of $10^{8.5-9.7}$ \msun\ in the fitting since we cannot probe typical star-forming 
galaxies at these bins due to the flux limit (\S\ref{s3.3}). Also, we remove the radio 
galaxy (M$_\star=10^{12.4}$ \msun). The thick frames in fig.~\ref{fig5} highlight the 
sample bins used in the curve fitting. 

\footnotetext[4]{\url{http://purl.com/net/mpfit}}

\begin{table}
\centering
\caption{Results of Schechter function fitting for the stellar mass distribution. The 
third and fourth columns indicate the normalisation factors between PKS~1138 and the 
general fields at the similar redshift range reported by \citet{Davidzon:2017} and 
\citet{Sobral:2014} at the stellar mass of $10^{9.7}$ \msun, respectively.}
\begin{tabular}{cccc}
\hline
log(M$_\star^\ast$/M$_\odot$) & log($\Phi_{\mathrm{M}_\star}^\ast$/Mpc$^{-3}$) & $\Phi_\mathrm{9.7}$/$\Phi_\mathrm{D17}$ & $\Phi_\mathrm{9.7}$/$\Phi_\mathrm{S14}$ \\
\hline
$11.726\pm0.756$ & $-3.097\pm0.432$ & 9.12 & 13.29 \\
\hline
\end{tabular}
\label{tab3}
\end{table}

The derived parameters of the curve fitting are given in table~\ref{tab3}. Due to the 
small sample size, we fix the power law slope to $\alpha=-1.5$, as used in 
\citet{Shimakawa:2018}, to minimise the fitting errors. We should note that our 
restricted sample sizes and large binning sizes would not be sufficient to determine 
the function parameters and even errors robustly. Indeed, it is known that there are 
non-negligible variations of derived M$_\star^\ast$ and $\Phi^\ast$ even if one 
employs much larger datasets, perhaps due to the cosmic variance and selection effects 
\citep{Ilbert:2013,Muzzin:2013,Sobral:2014,Davidzon:2017,Hayashi:2018b}. Although 
systematic comparisons are unfair because of such issues, the differences of the 
number densities between PKS~1138 and the general fields 
\citep{Sobral:2014,Davidzon:2017} at the stellar mass $\sim10^{10}$ \msun\ 
(fig.~\ref{fig5}) suggest that the PKS~1138 protocluster is an approximately ten times 
higher density region than the general field at a similar redshift. 

The figure also shows the stellar mass function of HAEs in the USS~1558 protocluster 
at $z=2.53$, and its group regions and intergroup regions \citep{Shimakawa:2018}. 
These functions are derived by the same procedure as this work, which enables a 
relatively fair comparison between two protoclusters at different redshifts. We find 
that the cut-off stellar mass of HAEs in PKS~1138 at $z=2.2$. 
Whilst the derived cut-off values would be overestimated since we may 
overestimate stellar masses of AGN host galaxies, this is consistent with that in the 
fragmented group regions (M$_\star^\ast\sim10^{11.5}$ \msun) in USS~1558 at $z=2.5$, 
within the error margins. Both have significantly higher characteristic stellar masses 
than those in the intergroup regions (M$_\star^\ast\sim10^{10.6}$ \msun) in USS~1558, 
meaning that PKS~1138 is associated with the larger number of more massive HAEs than 
lower density regions in USS~1558 at $z=2.53$. Also, this could suggest that PKS~1138 
is at a point where fragmented cores are about to consolidate into a massive cluster 
with a single core if we assume these two protoclusters are on the similar 
evolutionary track to massive clusters \citep{Shimakawa:2014}. Comparing these two 
protoclusters in the stellar mass function, the PKS~1138 region has a 1.5 times lower 
number density of HAEs relative to USS~1558.

\begin{figure}
\centering
\includegraphics[width=0.95\columnwidth]{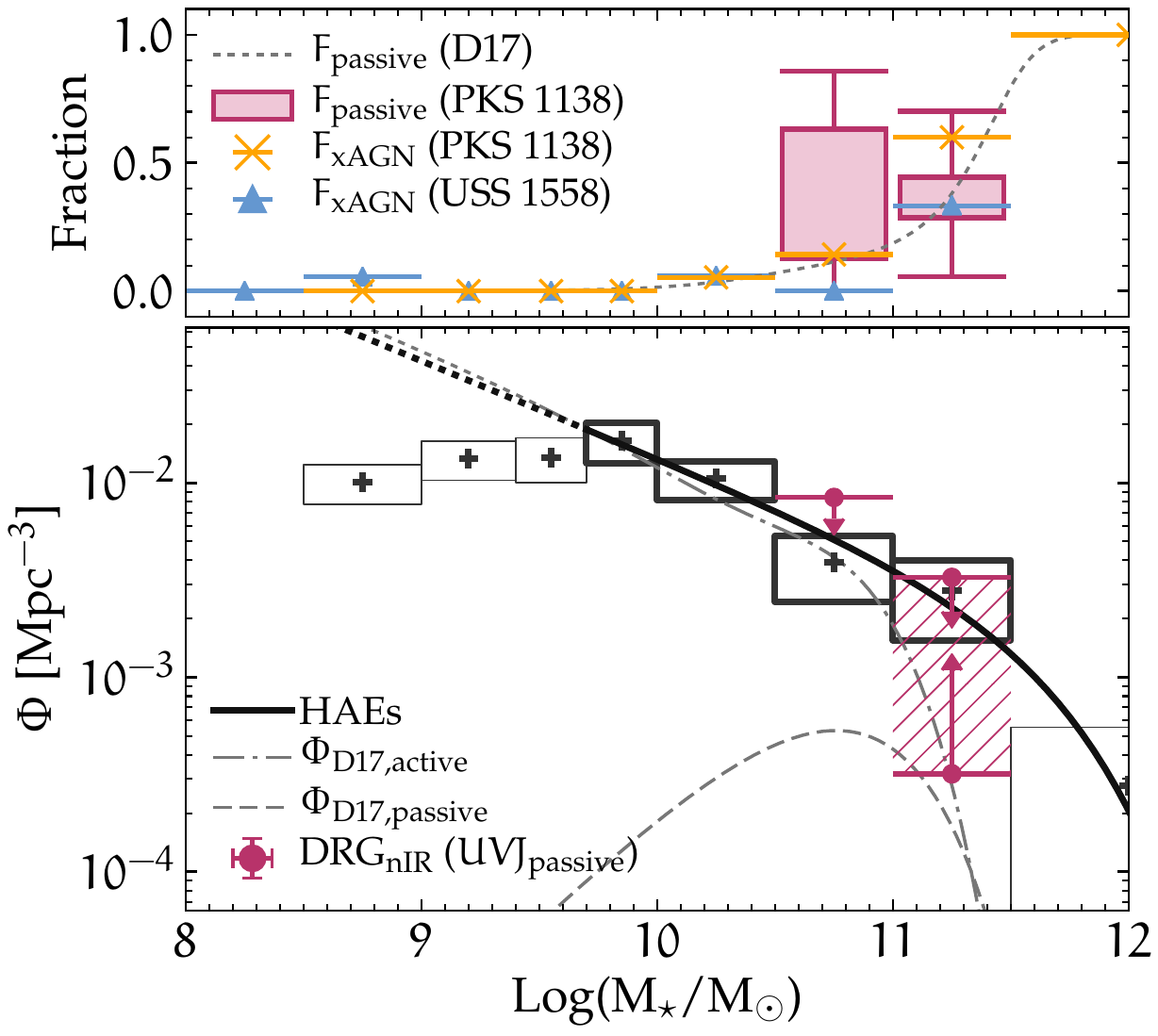}
\caption{{\it Lower panel:} Stellar mass function in the PKS~1138 protocluster region 
(same as in fig.~\ref{fig5}). This figure also plots the $1\sigma$ upper limits of the 
stellar mass function for passive galaxies at M$_\star>10^{10.5}$ \msun\ based on the 
DRG$_\mathrm{nIR}$ samples. We select DRG$_\mathrm{nIR}$ whose rest-frame $UVJ$ colours 
agree with passive galaxy populations within $1\sigma$ errors. The lower limits are 
constrained by three quiescent objects with spectroscopic (or spectrophotometric) 
confirmation by \citet{Tanaka:2013}. Grey dash-dotted and dashed curves are the stellar 
mass function of star-forming and passive galaxies in the general field at $z=$ 2--2.5 
\citep{Davidzon:2017}. 
{\it Upper panel:} Fraction of X-ray selected AGNs among HAEs (orange crosses) and 
passive galaxies (red hatched region) in PKS~1138 as a function of stellar mass. The 
latter is calculated based on the $1\sigma$ upper and lower limits shown in the lower 
panel, and error-bars include Poisson noise. Blue triangles indicate the X-ray AGN 
fraction among HAEs in USS~1558 at $z=2.53$ \citep{Macuga:2018}. The grey dotted line 
indicates the passive fraction in the general field at $z=$ 2--2.5 reported by 
\citet{Davidzon:2017}.}
\label{fig6}
\end{figure}

We also estimate the fraction of bright X-ray sources among HAEs in each stellar 
mass bin. Studying AGN activities across different stellar mass ranges is essential 
since AGNs are thought to have mass dependence and especially play critical roles in 
massive-end systems \citep{Ferrarese:2000,Kauffmann:2003,Matteo:2005}. We thus merely 
check the X-ray AGN fraction in our HAE samples and then find that more than 60 percent 
(4/6) of very massive HAEs host AGNs at M$_\star=10^{11-12.5}$ \msun\ (fig.~\ref{fig6}). 
Such a high AGN fraction may be even more enhanced once we get a better AGN 
identification tool (see discussion \S\ref{s4}). The figure also shows AGN fraction in 
the other protocluster area, USS~1558 at $z=2.53$ \citep{Macuga:2018}. This detection 
limit of $L_X=3\times10^{43}$ erg~s$^{-1}$ is deeper than that in PKS~1138. In 
USS~1558, despite that, there is no bright X-ray source in a higher density region 
within the protocluster except the radio galaxy that is the only X-ray source (1/3) in 
the highest stellar mass bin (M$_\star=10^{11}$--$10^{11.5}$ \msun). We summarise the 
AGN fraction in each stellar mass bin in table~\ref{tab4}. 

\begin{table}
\centering
\caption{X-ray AGN fractions amongst HAEs in PKS~1138 ($z=2.15$) and USS~1558 
($z=2.53$) at different stellar mass bins. One should note that these do not include 
the HAE candidates and the completeness correction.}
\begin{tabular}{lll}
\hline
log(M$_\star^\ast$/M$_\odot$) & PKS~1138 & USS~1558 \\
\hline
8.0--8.5 & --- & 0/5 (0\%) \\
8.5--9.0 & 0/3 (0\%) & 1/18 (6\%) \\
9.0--9.4 & 0/14 (0\%) & 0/21 (0\%) \\
9.4--9.7 & 0/10 (0\%) & 0/20 (0\%) \\
9.7--10.0 & 0/11 (0\%) & 0/17 (0\%) \\
10.0--10.5 & 1/17 (6\%) & 1/18 (6\%) \\
10.5--11.0 & 1/7 (14\%) & 0/5 (0\%) \\
11.0--11.5 & 3/5 (60\%) & 1$^\ast$/3 (33\%) \\
11.5--12.5 & 1$^\ast$/1 (100\%) & --- \\
\hline
\multicolumn{3}{l}{$^\ast$ Radio galaxies}\\
\end{tabular}
\label{tab4}
\end{table}

In addition, we tentatively constrain the quenching fraction in PKS~1138 by combining 
HAEs with DRG$_\mathrm{nIR}$ samples (\S\ref{s2.3.3}). Within the target area, there 
are 34 DRG$_\mathrm{nIR}$ sources that do not have flux excesses at narrow-band nor 
bright dust emission at MIPS/24$\mu$m band, which provide the upper limit of the 
distribution functions of passive galaxies in PKS~1138. We here employ only 
DRG$_\mathrm{nIR}$ that can be classified as passive populations on the rest-frame 
$UVJ$ plane within margins of $1\sigma$ errors. Also, three of them are 
spectroscopically (or spectrophotometrically) identified by \citet{Tanaka:2013}. They 
allow us to constrain the lower limit of number densities of passive objects. We derive 
their stellar mass from the SED-fitting by assuming the fixed redshift of 2.15 and then 
employ only the 23 DRG$_\mathrm{nIR}$ with the stellar mass greater than $10^{10.5}$ \msun\ 
in this analysis. 

Roughly expected number densities of passive galaxies and the passive fraction as a 
function of the stellar mass can be seen in fig.~\ref{fig6}. The fraction of passive 
galaxies in PKS~1138 is estimated to be $\sim36$ percent. One should note that, 
however, such a constraint becomes almost irrelevant once we include errors. When we 
compare the number density of our DRG$_\mathrm{nIR}$ samples at $K_s=$ 21--23 mag with 
that of DRG$_\mathrm{nIR}$ selected in the same way (\S2.3.3) in the COSMOS field 
\citep{Laigle:2016}, the excess factor of DRG$_\mathrm{nIR}$ in PKS~1138 is estimated 
to be $\sim1.7$ (which is consistent with the estimation by \citealt{Kodama:2007}). The 
upper limits of the number densities thus should be overestimated due to the foreground and 
background contaminants. A future deep NIR survey with instruments such as Keck/MOSFIRE 
and JWST/NIRSpec is required to obtain the passive fraction more reliably. 

\begin{figure}
\centering
\includegraphics[width=0.9\columnwidth]{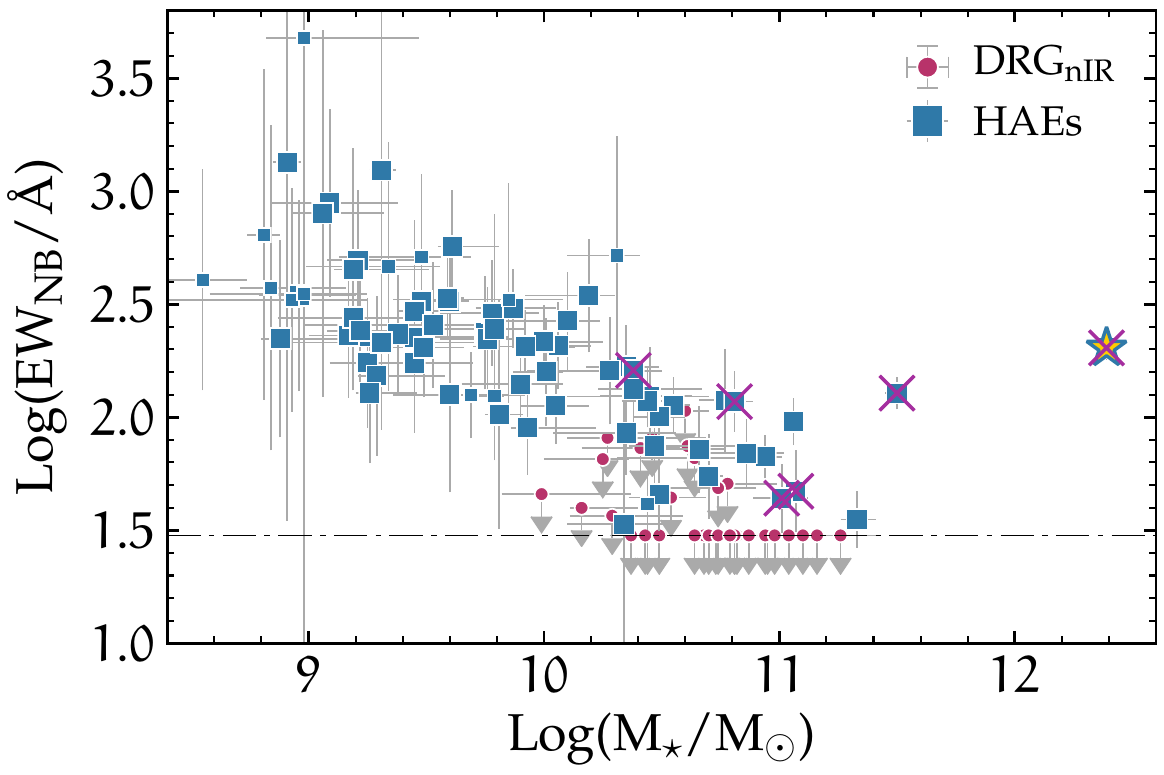}
\caption{EW of narrow-band flux (EW$_\mathrm{NB}$) against stellar mass. Blue large 
and small squares show HAEs and HAE candidates respectively (table~\ref{tab2}). The 
horizontal dash-dotted line indicates our EW selection limit that corresponds to the 
rest-frame EW$_\mathrm{NB}=30$ \AA\ at $z=2.15$. Purple crosses have X-ray 
counterparts. error-bars show $1\sigma$ uncertainties. We also plot upper limits of 
EW$_\mathrm{NB}$ for the DRG$_\mathrm{nIR}$ samples by red circles for a reference.}
\label{fig7}
\end{figure}

Regarding these trends, we caution about a potential issue in our narrow-band 
selection. Our HAE samples are limited either by the narrow-band flux 
($>3\times10^{-17}$ erg~s$^{-1}$cm$^{-2}$) or EW$_\mathrm{NB}$ ($>30$ \AA\ in the rest 
frame) depending on their narrow-band magnitude (fig.~\ref{fig2}). Figure~\ref{fig7} 
roughly explains how this selection bias may affect the different stellar mass ranges. 
Since the narrow-band magnitude correlates with the stellar mass, the flux limit is the 
primary bias in the narrow-band selection at the lower stellar mass regime, 
$\lesssim10^{10.4}$ \msun\ (i.e., eq.~\ref{eq1}). On the other hand, towards the 
massive end (M$_\star\gtrsim10^{10.4}$ \msun), the EW$_\mathrm{NB}$ limit ($>30$ \AA) 
as defined by eq.~\ref{eq2} drives the selection bias. Thus, we should note that our 
HAE selection is not fully equal to the selection of star-forming populations at the 
massive end, in the meaning that the selection is restricted by EW$_\mathrm{NB}$ that 
roughly corresponds to specific SFR instead of the narrow-band flux (i.e., SFR). Such 
an additional bias would require special attention when we regard our HAE samples as 
star-forming galaxies. For example, we may underestimate the number of star-forming 
galaxies at the massive end. Also, the AGN fraction could be overestimated since 
EW$_\mathrm{NB}$ could be enhanced by the flux contribution in both \ha\ and \nii\ 
lines from AGNs. For reference, \citet{Sobral:2016b} have found that AGNs contribute 
$\sim15$ percent of the total \ha\ luminosity density at any redshift up to $z=2.23$. 
We return to the discussion of the AGN fraction in \S4.

\citet{Shimakawa:2018} have derived the dust-corrected \ha\ luminosity function in the 
USS~1558 field, with tentative extinction correction for individual HAEs. However, we 
decide not to discuss the \ha\ luminosity function towards the PKS~1138 region since 
this protocluster is known to be associated with a large number of dusty starbursts 
\citep{Stevens:2003,Mayo:2012,Koyama:2013a,Valtchanov:2013,Rigby:2014,Dannerbauer:2014} 
and it is thus very challenging to properly estimate dust-corrected \ha\ luminosities 
for all HAEs on an equitable basis. If we apply dust correction in the same way as for 
HAEs in the USS~1558 field \citep{Shimakawa:2018}, we can derive 
$\log(L^\ast_\mathrm{H\alpha}$/erg~s$^{-1})=43.61\pm0.20$ and 
$\log(\Phi_{L_\mathrm{H\alpha}}^\ast$/Mpc$^{-3})=-2.38\pm0.18$, respectively.


\subsection{Spatial distribution and rest-frame colours}\label{s3.2}

\begin{figure*}
\centering
\includegraphics[width=0.95\textwidth]{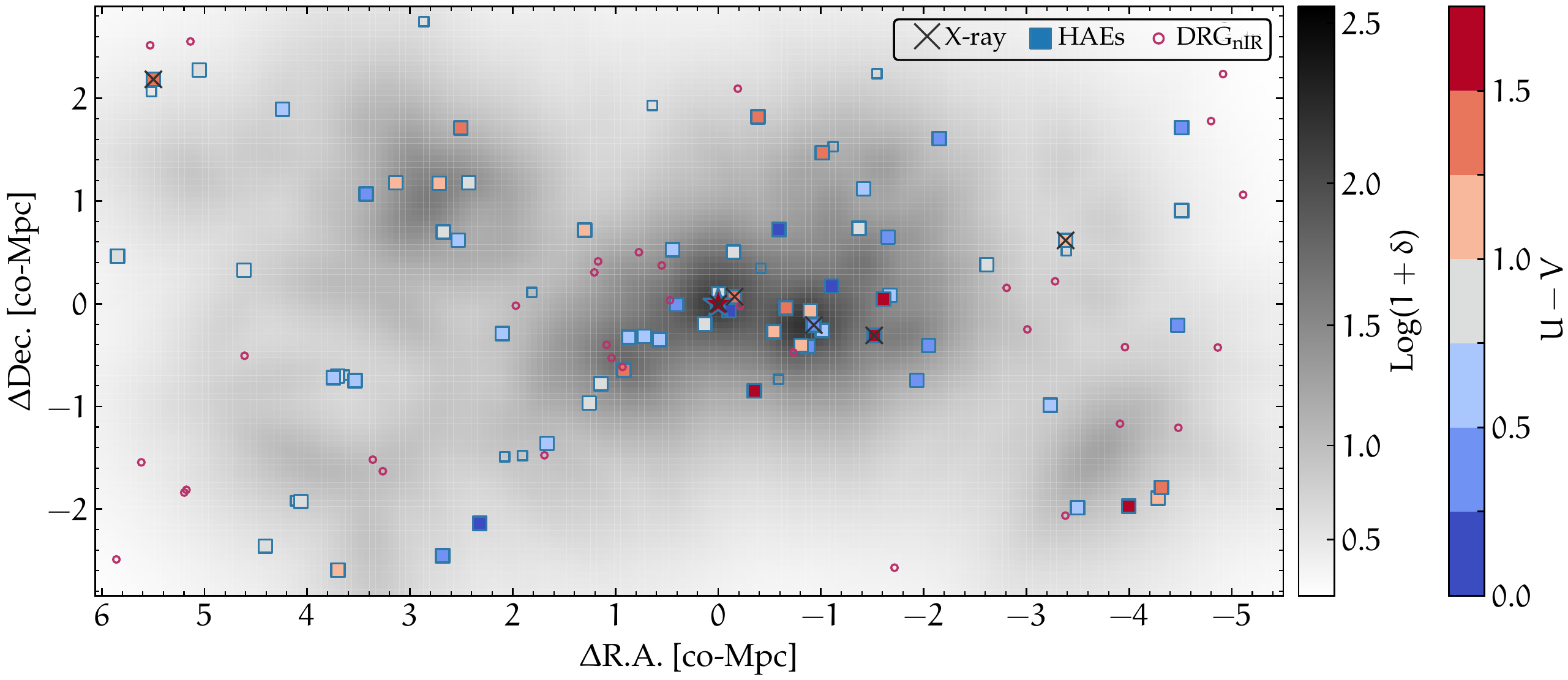}
\caption{Spatial distribution with respect to the Spiderweb galaxy as shown by the star 
symbol in the centre. Large filled and small open squares are HAEs and HAE candidates, 
respectively. Symbol colours of HAEs indicate their rest-frame $U-V$ colours. Red 
circles and black crosses indicate DRG$_\mathrm{nIR}$ and X-ray sources, respectively. 
The colour map in the background shows the excess of surface number densities based on the 
5th neighbour analysis.}
\label{fig8}
\end{figure*}

We show the positions of our updated HAE samples over the survey area in 
fig.~\ref{fig8}. The underlying colour in the figure indicates the excess of the 
number density ($\log(1+\delta)$), which is defined by the following equation, 
\begin{equation}
\delta = \frac{\Sigma_\mathrm{5th}-C\cdot\overline{\Sigma}_\mathrm{5th}}{C\cdot\overline{\Sigma}_\mathrm{5th}}
\label{eq7}
\end{equation}
where $\Sigma_\mathrm{Nth}$ is the surface density in physical (ph-) Mpc$^{-2}$ 
including $N$ HAEs therein. We here adopt $N=5$. $C$ is the scaling factor which 
normalises the mean number density over the entire protocluster field 
($\overline{\Sigma}_\mathrm{5th}=13.5$ ph-Mpc$^{-2}$) to the typical density in 
the general field. We tentatively apply $C=0.1$ inferred from the ten times higher 
number density of the stellar mass function at M$_\star\sim10^{10}$ \msun\ relative to 
those in the general fields (\S\ref{s3.1}). 

The PKS~1138 protocluster is associated with a very dense region of HAEs around the 
Spiderweb galaxy as firstly reported by \citet{Kurk:2004a}: it contains more than 100 
times as many galaxies as seen in the random field in the local scale (see also 
\citealt{Kuiper:2011}). We find $>10$ new \ha\ detections associated with this massive 
structure. Also, four out of six HAEs with X-ray detections \citep{Pentericci:2002} 
are positioned within or close to the central system. PKS~1138 is also known to have 
filamentary structures on the east side \citep{Croft:2005,Koyama:2013a} which are 
aligned along the line of sight as well \citep{Shimakawa:2014}. The most compact group 
in this region can be seen at four co-Mpc away eastward from the radio galaxy. This 
compact group involves four HAEs (\#25,26,27,29) within only 60 ph-kpc distance and 
has a $3.6\sigma$ source detection ($5.0\pm1.4$ mJy) at LABOCA 870 $\mu$m (\#DKB12 in 
\citealt{Dannerbauer:2014}). The peak density of passive galaxy candidates selected as 
DRGs is slightly shifted towards the east direction, though more spectroscopic 
identifications are needed to confirm this sub-structure.

\begin{figure}
\centering
\includegraphics[width=0.9\columnwidth]{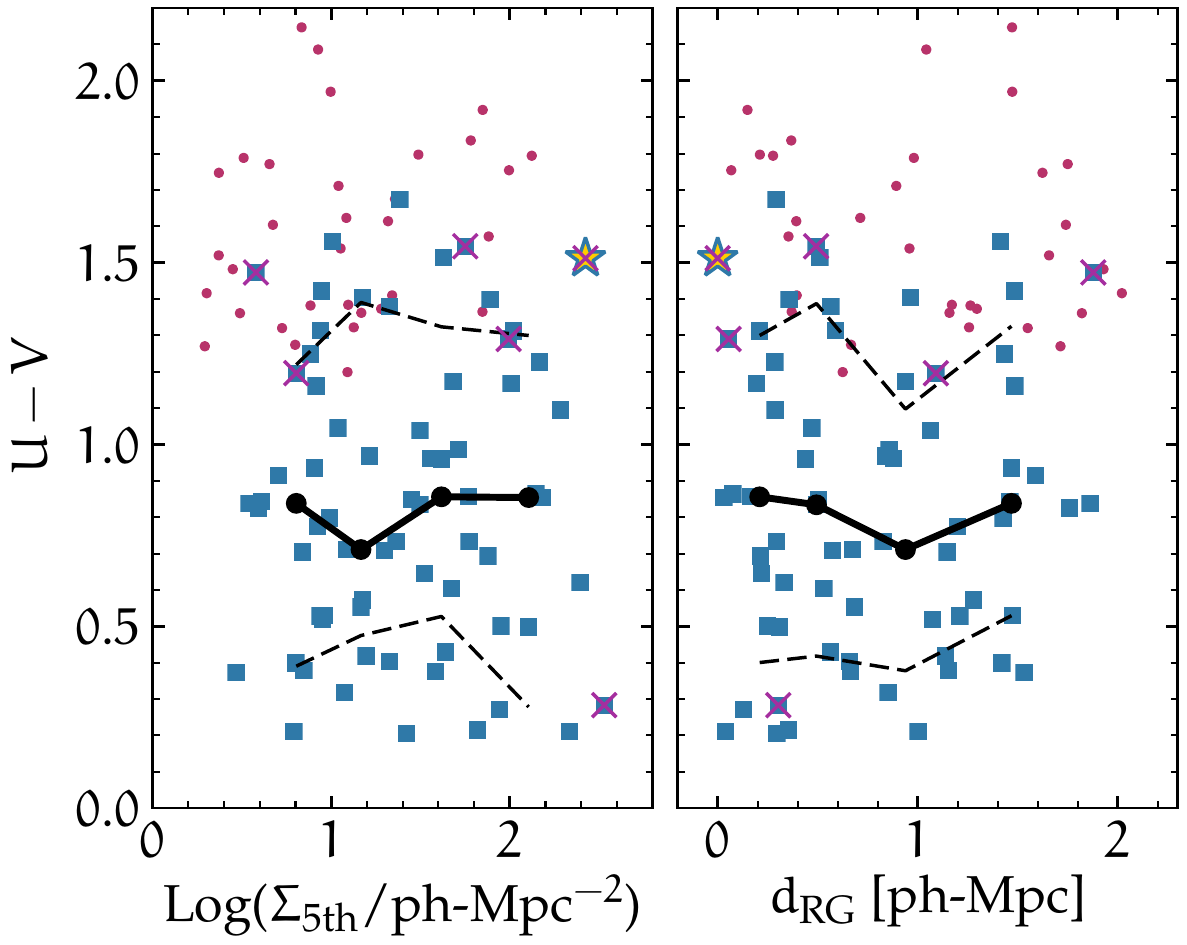}
\caption{Rest-frame $U-V$ colours as a function of surface number densities of HAEs 
(from the left, $\Sigma_\mathrm{5th}$ ph-Mpc$^{-2}$ and distance from the Spiderweb 
radio galaxy, $d_\mathrm{RG}$ ph-Mpc). The symbols are the same as in fig.~\ref{fig7}. 
Solid and dashed lines indicate median values and 68th percentiles distributions of 
rest-frame $U-V$ colour of HAEs with respect to each axis.} 
\label{fig9}
\end{figure}

We then estimate the rest-frame $UVJ$ colours, and associated errors, of HAEs using 
the EAZY code \citep{Brammer:2008,Brammer:2011}, and then investigate colour 
dependence on the local scale. The rest $U$, $V$, $J$ photometries roughly correspond 
to $Y$, $H$, and 3.6 $\mu$m bands, respectively. 

Figure~\ref{fig9} shows the surface number densities vs. the rest-frame $U-V$ colours 
of HAEs. We explore the colour dependence on local environments with different density 
measurements, i.e., the surface densities including 5th neighbours, and distance from 
the radio galaxy ($d_\mathrm{RG}$). We find no clear correlation between $U-V$ colours 
and local densities of HAEs within the protocluster, which is unchanged when we use 
$\Sigma_\mathrm{3th}$ or $\Sigma_\mathrm{10th}$ for the density measurement. These 
results are inconsistent with the concentration of redder HAEs towards the 
protocluster centre as reported by \citet{Koyama:2013a}. However, \citet{Koyama:2013a} 
investigated the colour dependence of HAEs in an area twice as large as the survey 
field of MDCS, and their results are enhanced by the absence of red HAEs in these 
outer regions. We, therefore, conclude that the inconsistency between this work and 
\citet{Koyama:2013a} is due to the insufficiently large survey area in this work to 
confirm the finding of \citet{Koyama:2013a}. 

\begin{figure}
\centering
\includegraphics[width=0.9\columnwidth]{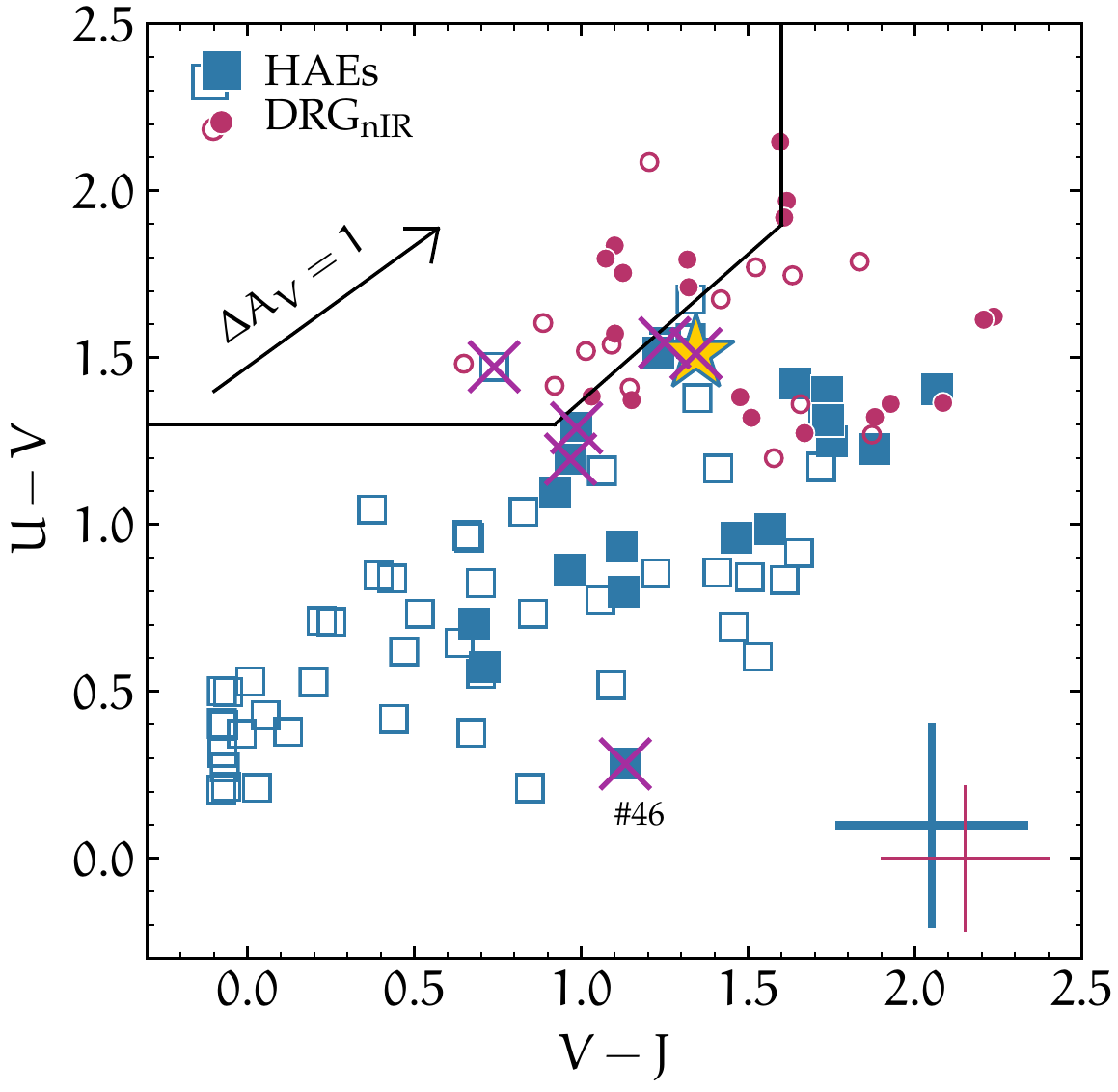}
\caption{Rest-frame $UVJ$ colours of our HAE and DRG$_\mathrm{nIR}$ samples in the 
protocluster region. Blue squares and red circles are HAEs and DRG$_\mathrm{nIR}$, 
respectively. Filled symbols indicate a $>2\sigma$ detection in either 3.6 or 4.5 
$\mu$m, or both. We do not have sufficient IRAC photometry for open symbols, and those 
$VJ$ colours are not reliable. Blue bold and red thin crosses on the lower right 
indicate typical errors of IRAC-detected HAEs and DRG$_\mathrm{nIR}$, respectively. 
According to the EAZY code, the typical uncertainty of the objects without IRAC 
detections ($>2\sigma$) is $\sim0.6$ dex. The black arrow corresponds to the colour 
shift by dust reddening ($\Delta A_V=1$) along the \citet{Calzetti:2000} extinction 
law. The black solid line indicates the quiescent/star-forming classification line 
from \citet{Williams:2009}.}
\label{fig10}
\end{figure}

We then characterise HAEs on the rest-frame $UVJ$ plane (fig.~\ref{fig10}). The 
shallow depths of the IRAC bands ($m_{3\sigma}\sim21.5$ AB), mean that only 32 percent 
of the entire HAE sample are detected at IRAC bands at a more than two sigma 
confidence level. These IRAC detected HAEs are shown by the filled symbols in 
fig.~\ref{fig10}. Typical errors are $\Delta(U-V)=0.31$ dex and $\Delta(V-J)=0.29$ 
dex, respectively. The remainders are indicated by open symbols and have rest $J$-band 
magnitudes estimated from the extrapolated SED spectra. Uncertainties of rest $V-J$ 
colours in these non-IRAC detections would be $\sim0.6$ dex according to the EAZY 
code. 

As a result, we find that rest-frame $UVJ$ colours of HAEs agree with those of the 
star-forming population \citep{Williams:2009,Moresco:2013,Straatman:2016,Fang:2018} 
within the margin of error. Despite the significant uncertainties of individual 
colours, HAEs hosting bright X-ray AGNs tend to have redder rest-frame $U-V$ colours, 
which agree with the findings by \citet{Krishnan:2017}. More interestingly, we see that 
HAEs with X-ray emissions (\#40,58,68,73,95) are preferentially located near the edge 
of the quiescent population. The outlier lying at the bottom on the $UVJ$ plane (\#46) 
is known to be an AGN (\#6 in \citealt{Pentericci:2002} and \#215 in 
\citealt{Kurk:2004b}), with very broad \ha\ line emission identified by near-infrared 
spectroscopy with VLT/ISAAC \citep{Kurk:2004b}. The colour trend suggests that HAEs 
hosting X-ray AGNs could be in the transition stage from dusty star-forming galaxies 
to passive populations, i.e., in the post-starburst phase. 
Another causal factor is the effect of the nuclear emission. AGN 
contribution would redden more rest-frame $V-J$ colours than $U-V$ colours, and thus 
this bias may rather weaken the colour discrepancy between HAEs and X-ray HAEs, 
though a more detailed analysis is needed.

\subsection{Star-forming main sequence}\label{s3.3}

We estimate the SFR$_\mathrm{H\alpha}$ of HAEs from the observed \ha\ luminosities 
(\S\ref{s2.4.2}) using the same method as \citet{Shimakawa:2018}. We use the 
\citet{Kennicutt:1998} prescription on the assumption of \citet{Chabrier:2003} IMF (a 
factor of 1.7 reduces SFR from the standard \citealt{Kennicutt:1998} calibration). We 
then correct for dust extinction of the \ha\ line based on the SED-inferred stellar 
extinction ($A_V$) from the {\sc fast} code (\S\ref{s2.4.1}) and the 
\citet{Calzetti:2000} extinction law. Throughout the analysis, we used the additional 
assumption of E($B-V$)$_\mathrm{stellar}$ = E($B-V$)$_\mathrm{nebular}$. Such a 
hypothesis is relatively reasonable for typical star-forming galaxies at $z\sim2$ 
\citep{Reddy:2015}. However, \citet{Reddy:2015} and \citet{Price:2014} also note that 
the ratio of nebular extinction to stellar extinction depends on galaxy properties, 
especially SFR; our assumption, therefore, would lead to underestimating the SFRs of 
active and dusty star-forming objects. Previous infrared studies also suggest that the 
dust extinction law for individual galaxies varies depending on their IR luminosities 
and dust geometries \citep{Reddy:2006,Reddy:2010,Casey:2014,Narayanan:2018}. 

\begin{figure}
\centering
\includegraphics[width=0.9\columnwidth]{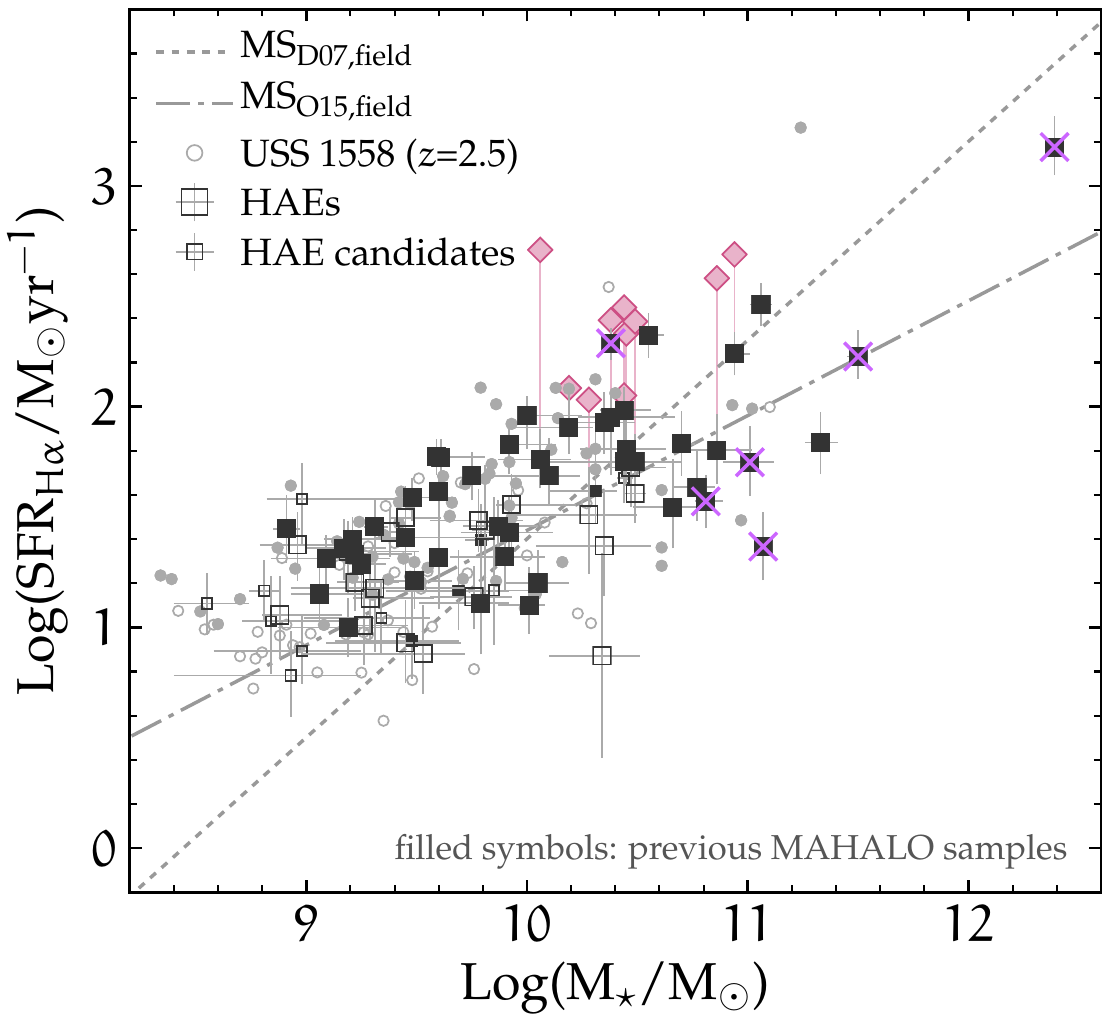}
\caption{Stellar mass versus dust-corrected SFR$_\mathrm{H\alpha}$ of HAEs. Black 
filled and open squares indicate HAEs and HAE candidates respectively 
(table~\ref{tab2}). The error-bars correspond to $1\sigma$ errors. Grey circles show 
HAEs in the USS~1558 protocluster at $z=2.53$ reported by \citet{Shimakawa:2018} which 
estimate SFR$_\mathrm{H\alpha}$ in the same way. Purple crosses represent HAEs 
overlapping with the Chandra X-ray sources whose observed \ha\ fluxes may be 
overestimated since we do not consider the flux contribution from AGNs. Grey 
dot-dashed and dashed lines are the star-forming main sequence of HAEs at $z=2.2$ in 
the general field \citep{Oteo:2015} and star-forming galaxies at $z\sim2$ reported by 
\citet{Daddi:2007a}, respectively. Pink diamonds show SFRs derived from UV and IR 
luminosities for MIPS/24$\mu$m detected HAEs (see text).}
\label{fig11}
\end{figure}

Indeed, we confirm that SFR$_\mathrm{UV+IR}$ of dusty HAE sources with Spitzer/MIPS 
24 $\mu$m detections is significantly higher than those dust-corrected 
SFR$_\mathrm{H\alpha}$ with our tentative extinction correction (fig.~\ref{fig11}). 
Their IR luminosities are derived from the flux densities at 24 $\mu$m by following 
the \citet{Wuyts:2008} conversion prescription and then obtaining their SFRs via 
SFR$_\mathrm{UV+IR}$ ($=1.09\times10^{-10}(L_\mathrm{IR}+3.3\nu L_{\nu,2800}$), 
\citealt{Bell:2005}). We only measure IR luminosities of 10 HAEs without X-ray 
detections (\#5,9,13,14,22,27,61,71,80,93). Two of them (\#27,93 but note \#27 is 
blended with \#26,29 in the MIPS/24$\mu$m image) are associated with 870 $\mu$m LABOCA 
sources (DKB12 and DKB15 in \citealt{Dannerbauer:2014}). According to the difference 
between dust-corrected SFR$_\mathrm{H\alpha}$ and SFR$_\mathrm{UV+IR}$, we expect to 
underestimate SFR$_\mathrm{H\alpha}$ by a factor of four on average for these dusty 
HAEs. Such a large mismatch is the reason why we do not analyse the \ha\ luminosity function in this paper, as discussed in the previous subsection. 

The new deeper \ha\ data succeed in finding relatively fainter HAEs than those found 
by the previous MAHALO-Subaru survey \citep{Koyama:2013a}. As shown in fig.~\ref{fig11}, when 
we apply the same method for extinction correction of the \ha\ line in HAEs in PKS~1138 
as used for those in USS~1558 \citep{Shimakawa:2018}, HAEs in PKS~1138 and USS~1558 
seem to follow the same stellar mass -- SFR$_\mathrm{H\alpha}$ relation. We find no 
clear trend in star formation with environment, which is in agreement with past studies 
in the same field \citep{Koyama:2013a,Koyama:2013b}. However, finding no environmental 
dependence on star formation in PKS~1138 is in contrast to USS~1558 where the HAEs in 
the group regions are more actively star-forming than those in the intergroup regions 
\citep{Shimakawa:2018}. Moreover, we find that three of the X-ray detected HAEs tend to 
be positioned below the main sequence, although these SFR$_\mathrm{H\alpha}$ would be 
contaminated by additional \ha\ emission from AGNs. This is consistent with their 
post-starburst like rest-$UVJ$ colours (\S\ref{s3.2}), though we should keep in mind 
that our dust corrections mass estimates of X-ray hosts have 
substantial uncertainties. One source showing the excess of star formation among HAEs 
with X-ray emission is HAE-\#46 (M$_\star=10^{10.4}$ \msun) that has a blue $U-V$ 
colour and \citet{Kurk:2004b} have identified that this HAE (\#215 in their literature) 
has a broad \ha\ line. 

Four HAEs (\#42,48,54,56) are confirmed by ALMA CO($3-2$) observations (Tadaki et al. 
in preparation). We cannot use these data to derive their dust-corrected SFRs more 
robustly as they are clearly blended with other sources in the MIPS/24 $\mu$m band. 
However, \#48,54 are more like dusty starbursts according to their rest-frame $UVJ$ 
colours (appendix~\ref{a3}). 

Our deeper \ha\ imaging increases the number of HAE sources by 40--50 percent compared 
to our previous narrow-band imaging \citep{Koyama:2013a}. The mean narrow-band flux of 
newly identified HAEs is 2.5 times fainter than the mean flux of the entire sample. We 
discover eight HAEs at a stellar mass lower than $10^{9}$ \msun\ similar to those  
found in USS~1558 at $z=2.5$ \citep{Hayashi:2016}. The larger number of lower-mass 
HAEs in USS~1558 (fig.~\ref{fig11}) should be due to its 0.5 mag deeper $K_s$ data than 
that in PKS~1138. We defer the detailed analyses of these samples to future work since 
it remains unclear whether they are truly HAEs associated with the PKS~1138 
protocluster, due to a lack of firm detections at multi photometric bands. Future 
follow-up deep spectroscopy can provide us with both robust confirmations of these 
faint sources, and improved completeness when we investigate gas-phase metallicities 
of low-mass HAEs as compared to the previous spectroscopic analyses by 
\citet{Shimakawa:2015}.


\subsection{The Spiderweb nebula}\label{s3.4}

The previous \ha\ line observations reported that the Spiderweb radio galaxy is 
associated with an extended \ha\ nebula over a few ten ph-kpc in radius 
\citep{Kurk:2002,Nesvadba:2006,Kuiper:2011,Koyama:2013a}. \citet{Kurk:2002} have 
derived \lya/\ha\ line ratios in different regions within the \ha\ (\lya) nebula. 
These vary between 0.015 (at the nucleus) and 7.6 (at the position blending with the 
radio jet). \citet{Nesvadba:2006} have studied the spatial distribution and kinematics 
of the optical emission lines in detail by using the near-infrared integral field 
spectrograph (VLT/SPIFFI). Their emission line analyses suggest that the Spiderweb 
nebula has electron density $n_e\sim388_{-148}^{+182}$ cm$^{-3}$, broad-line emission 
with FWHM$>2000$ km~s$^{-1}$, typical \nii/\ha\ line flux ratio of $\sim1$, and the 
total observed \ha\ luminosity of $L_\mathrm{H\alpha}=(14.8\pm1.2)\times10^{43}$ 
ergs~s$^{-1}$. They roughly estimate the least total ionised gas mass of 
M$_\mathrm{H\textsc{ii}}=$ (2.3--6.5)$\times10^9$ \msun. Such a relatively high \ha\ 
emission brightness cannot be explained by shock heating alone \citep{Dopita:1996}, so 
\citet{Nesvadba:2006} conclude that photoionisation from the radio galaxy plays a 
dominant role in the emission line properties of this large \ha\ nebula (see also 
\citealt{Martin:2003,Nesvadba:2008,Nesvadba:2017}). 

\begin{figure*}
\centering
\includegraphics[width=0.95\textwidth]{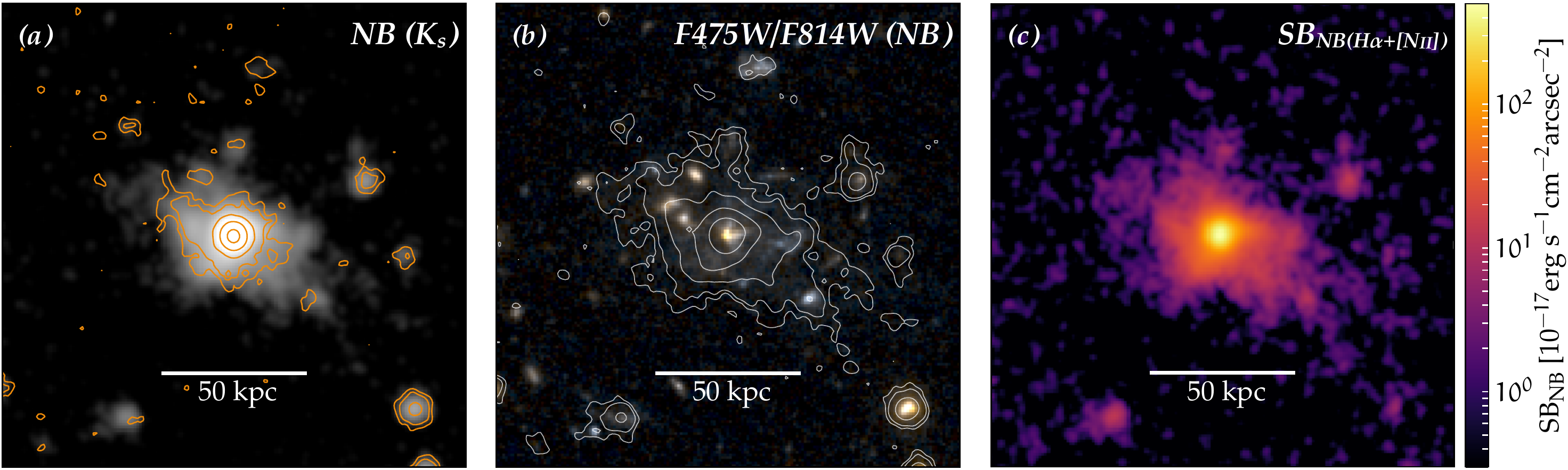}
\caption{Multi-faces of the Spiderweb galaxy (20 arcsec on a side in each image).
(a) The left panel shows NB$_{2071}$ image (grey scale) and line-subtracted $K_s$ 
image (orange contour). The first contour is 1.5 sigma background rms. 
(b) The middle panel is an RGB image from F475W and F814W photometry with HST. White 
contours are based on the NB$_{2071}$ image, and the first contour corresponds to 1.5 
rms in the background. 
(c) The right panel represents the surface brightness of narrow-band flux 
(SB$_\mathrm{NB}$) where the continuum is subtracted (eq.~\ref{eq3}). 1 sigma error is 
$3.2\times10^{-17}$ erg~s$^{-1}$cm$^{-2}$arcsec$^{-2}$ for a given pixel$^2$ area.}
\label{fig12}
\end{figure*}

Our deep \ha\ narrow-band imaging can trace extended \ha\ (+\nii) emission over the 
wider FoV, down to $4.8\times10^{-17}$ erg~s$^{-1}$cm$^{-2}$arcsec$^{-2}$ above the 
1.5 sigma level in pixel scale. As a result, we succeed in identifying the large \ha\ 
nebula structure that has a maximum projected extent of $\sim100$ ph-kpc 
(fig.~\ref{fig12}). Such an enormous \ha\ structure has never been discovered before 
at $z>2$. The \ha\ nebula extends in the NE--SW direction, which is broadly consistent 
with the inclination of structures in X-ray and CO($1-0$) emission found by 
\citet{Carilli:2002} and \citet{Emonts:2016}, respectively. 

When we adopt the conservative assumption of \nii/\ha=1 in the same manner as 
\citet{Nesvadba:2006}, we obtain the total observed \ha\ luminosity 
$L_\mathrm{H\alpha}=(1.35\pm0.15)\times10^{44}$ ergs~s$^{-1}$ over 50 ph-kpc in 
radius. This value is consistent with the past result by \citet{Nesvadba:2006}. We 
should note that the obtained \ha\ luminosity is a lower limit since dust correction is 
not implemented and the bandwidth of the narrow-band filter does not fully cover the 
broad-line emission as identified by \citet{Nesvadba:2006}. These complicated issues 
prevent us from exploring this unique nebula in more detail. We experimentally carry 
out some further analyses with these caveats by comparing with HST/ACS images, which 
are shown in appendix~\ref{a4}. 

Finally, we inspected the other HAEs in PKS~1138 searching for significantly extended 
\ha\ structures based on the median stacking with the {\sc iraf} scripts. Our stacking 
procedure reaches down to $6.4\times10^{-18}$ erg~s$^{-1}$cm$^{-2}$arcsec$^{-2}$ with 
a two sigma confidence level. However, we do not identify such diffuse nebula beyond a 
radii of 10 ph-kpc in the other HAEs, even after stacking all HAE samples.


\section{Discussion}\label{s4}

High-redshift protoclusters provide us with a great opportunity to unveil the 
formation histories of massive galaxy clusters and their member galaxies in the local 
Universe. As claimed by past studies 
\citep{Kurk:2004a,Kodama:2007,Doherty:2010,Hatch:2011,Galametz:2012,Koyama:2013a,Tanaka:2013}, 
PKS~1138 at $z=2.2$ is considered to be one of the most massive protoclusters at 
$z\sim2$, with a significant number excess ($\delta\gtrsim10$) of massive red 
star-forming and passive galaxies over the MOIRCS FoV relative to the general field at 
a similar redshift. This suggests that the massive galaxies in PKS~1138 are in the 
maturing phase, and such red galaxies would provide us with direct insights into the 
quenching processes of bright red sequence objects seen in the present-day clusters of 
galaxies. 

With such expectations, the original motivation of this work was to characterise these 
massive HAEs. We especially focus on X-ray fraction and the rest-frame $UVJ$ colours of 
HAEs based on the Chandra X-ray data and multi-band images, both of which mostly cover 
the survey area. We first derive the stellar mass function and confirm the number 
excess of massive HAEs seen in the past studies 
\citep{Hatch:2011,Koyama:2013a,Koyama:2013b}. The characteristic stellar mass is 
comparable to that in dense group cores seen in the USS~1558 protocluster 
\citep{Shimakawa:2018}. However, PKS~1138 is associated with a larger number of massive 
red HAEs than USS~1558 \citep{Kodama:2007,Galametz:2012}, suggesting that PKS~1138 is a 
more developed system. In addition, considering the fact that the PKS~1138 protocluster 
hosts at least three passive galaxies in such a small field 
(\citealt{Doherty:2010,Tanaka:2013}, see also \citealt{Zirm:2008}), PKS~1138 is ideally 
placed to see the rapid transition from massive dusty star-forming galaxies to 
quiescent galaxies. 

We find that four out of the six very massive HAEs (M$_\star=10^{11-12.5}$ \msun), 
including the Spiderweb galaxy, host bright X-ray sources ($L_X\gtrsim10^{44}$ 
erg~s$^{-1}$, see also \citealt{Pentericci:2002}). In general fields at the similar 
redshift, \citet{Sobral:2016b} and \citet{Matthee:2017} have shown that the AGN 
fraction significantly increases with increase in \ha\ luminosities of HAEs, and 
reaches $\sim100$ percent at the bright end ($>10^{43.5}$ 
erg~s$^{-1}$ without dust correction). Similarly, we confirm that the top two HAEs 
with high observed narrow-band luminosities ($=10^{43-43.5}$ erg~s$^{-1}$ without 
dust correction) and the Spiderweb radio galaxy ($=10^{44}$ erg~s$^{-1}$) host X-ray 
counterparts. The fraction is about two times higher than those in general fields 
reported by them, although our sample size is quite small. Such a relatively fair 
comparison seems to strength our finding of the enhancement of AGN fraction in the 
protocluster region.
Intriguingly, they tend to have unique rest-frame $UVJ$ colours corresponding to the 
post-starburst sequence between star-forming galaxies and passive populations 
\citep{Williams:2009,Whitaker:2011}. 
Thus, AGNs may play an important role in suppressing the star formation and the 
build-up of massive quiescent galaxies since these massive HAEs are more likely to grow 
into the bright-end red sequence systems seen in the massive galaxy clusters in the 
local Universe. AGN feedback is indeed considered to be an important mechanism to 
suppress the star formation of massive systems as pointed out by previous work (e.g., 
\citealt{Springel:2005,Croton:2006,Bower:2006,Somerville:2008,Voort:2011b,Page:2012,Kormendy:2013,Cicone:2014,Genzel:2014b}).
If a significant number of massive forming galaxies are subject to the influence of 
such strong feedback, we could explain the rapidly declining star formation around the 
distant cluster centres since $z\sim2$ 
\citep{Smail:2014,Shimakawa:2014,Clements:2014,Kato:2016}. 

We should note that the derived AGN fraction among HAEs is the minimum fraction 
inferred from the Chandra X-ray data. Our previous spectroscopic analyses 
\citep{Shimakawa:2015} indicate that other three sources (\#14,30,54) might also be 
AGNs because of their high \nii/\ha\ line ratios ($>0.5$), although the spectral data 
are obtained with low spectral resolution (R=513) and the derived line ratios have 
substantial uncertainties ($\sim0.2$ dex). If we assume that these HAEs also host AGNs, 
the AGN fractions increase to 43 and 80 percent at the stellar mass bins of 
$10^{10.5-11}$ and $10^{11-11.5}$ \msun\ respectively. 
Such a large number of AGN host galaxies may affect the derivation of the stellar mass 
function, since their stellar mass estimates may be overestimated due to the flux 
contamination from the nuclear emission. In the meanwhile, the surprisingly high AGN fraction 
implies that massive star-forming galaxies in PKS~1138 may mostly host AGNs and could 
have a considerable impact on the current galaxy evolution paradigm. 

The most energetic source present is the Spiderweb radio galaxy. We identified the 
enormous \ha\ structure extending over 100 ph-kpc associated with this massive  
system. The spatial extent broadly agrees with previous findings which have reported 
an extended component in the \lya\ line \citep{Gopal:2000,Kurk:2002}, X-ray 
\citep{Carilli:2002}, UV \citep{Pentericci:1998,Hatch:2008}, and the CO($1-0$) line 
\citep{Emonts:2016}. The Spiderweb galaxy is also known to be a composite of active 
star formation and AGN \citep{Ogle:2012,Drouart:2014}. The starburst component has 
SFR of $\sim1400$ \msun~yr$^{-1}$ \citep{Seymour:2012,Rawlings:2013,Dannerbauer:2014}, 
supported by a rich gas reservoir of $\sim6\times10^{10}$ \msun\ 
\citep{Emonts:2013,Emonts:2016,Gullberg:2016}. This monster galaxy is expected to grow 
into a brightest cluster galaxy, as seen in the local Universe \citep{Hatch:2009}. The 
large \ha\ nebula could show the occurrence of a pre-heating event in the 
proto-intercluster medium \citep{Babul:2002,Dubois:2011,Dubois:2012,Valentino:2016}, 
though there is still much debate about heating mechanisms of the intercluster medium 
at high redshifts \citep{McNamara:2007,Kravtsov:2012}. 

Lastly, the very high X-ray fraction in massive HAEs cautions the possibility that the 
narrow-band HAE selection may overestimate an AGN fraction, especially at the massive 
end. For example, \citet{Sobral:2016b} have shown that AGNs typically contribute 15 
percent of the total \ha\ luminosity. Moreover, AGNs are expected to enhance \nii\ line 
flux \citep{Baldwin:1981,Veilleux:1987} and this can additionally increase the 
narrow-band flux and EW$_\mathrm{NB}$. For instance, for a given \ha\ luminosity, an 
increase of log \nii/\ha\ ratio from $-0.5$ to $0$ raises EW$_\mathrm{NB}$ by 0.18 dex. 
These contaminations may cause the serious selection bias in the sense that the 
narrow-band selection preferentially detect AGN host HAEs. In particular, our current 
analyses cannot rule out the possibility that we are missing massive dusty starburst 
populations which would have faint observed \ha\ luminosities and low EW$_\mathrm{NB}$ 
due to heavy dust obscuration. Even if we miss a few dust-obscured objects, this factor 
can substantially decrease the bright-end AGN fraction. A wide-field spectroscopic 
search at the IR to radio regime with e.g., ALMA and JWST is highly desirable to 
clarify this selection issue.



\section{Summary}\label{s5}

In this second instalment of the MDCS campaign, we investigate HAEs associated with 
one of the most studied protoclusters, PKS~1138 at $z=2.2$. Using the advanced dataset 
we construct samples of HAEs associated with the Spiderweb protocluster; these consist 
of 68 HAEs (36 confirmed with spec-$z$ and 32 selected by $Bz'K_s$ colour) and 13 HAE 
candidates. 17 and 9 objects amongst them are newly-discovered HAEs and HAE candidates 
by this work, respectively. The online catalogue (appendix~\ref{a3}) lists coordinates, 
confirmation status, physical properties, and rest-frame $UVJ$ colours of the HAE 
samples. The major findings are summarised as follows. 

\begin{description}

\item[---]
We investigate the stellar mass distribution function of HAEs in PKS~1138, including 
the completeness correction derived using a Monte Carlo simulation. We then identify 
the high cut-off stellar mass of log(M$_\star$/\msun) $=11.73\pm0.76$ and find that 
the number density is about ten times higher over the survey area than that in the 
blank field. On the other hand, we find that at least four out of six very massive HAEs 
have X-ray counterparts, and those stellar masses may be significantly overestimated. 
This uncertainty would especially affect the derivation of the characteristic stellar mass. 
The Spiderweb nebula is unique and shows an extensive structure over 100 ph-kpc in 
the \ha\ emission line, as well as in other tracers such as \lya, X-ray, UV, and CO($1-0$) 
lines. We also tentatively obtain a passive fraction of 36 percent by combining the HAE 
sample with DRG$_\mathrm{nIR}$, although the measurement has a substantial error. 

\item[---]
We investigate the rest-frame $UVJ$ colours of HAEs and their environmental dependence 
on the local scale. Given the limited survey field, we do not see any clear correlation 
between colours and local overdensities. On the other hand, HAEs with luminous X-ray 
emission tend to have post-starburst like colours, implying that bright AGNs may play 
an important role in quenching active star formation in these sources. Because the 
Spiderweb protocluster is also associated with a large number of dusty starbursts and 
passive galaxies, many massive galaxies in PKS~1138 seem to be undergoing a rapid 
transition from the active star formation phase into the quiescent mode. 

\end{description}

\begin{figure*}
\centering
\includegraphics[width=0.9\textwidth]{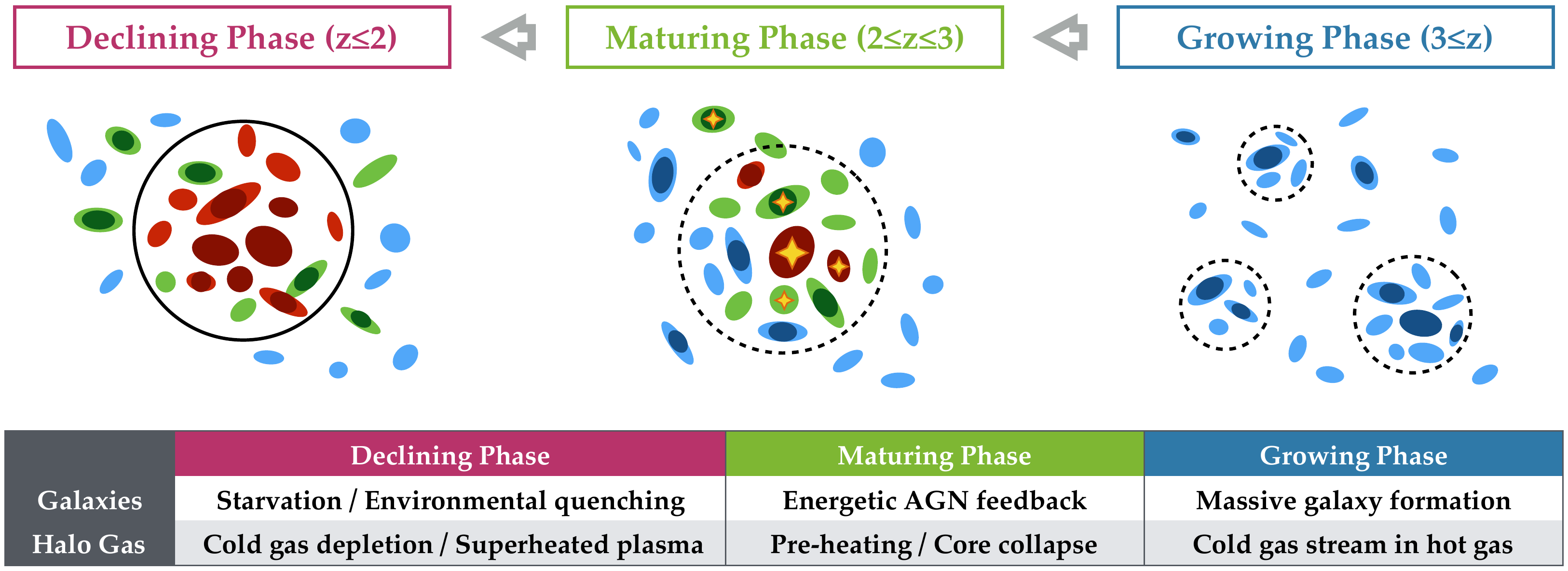}
\caption{A speculative schematic of formation and evolution histories of galaxy 
clusters and galaxies therein, based on our findings through the MDCS. We tentatively 
assume that PKS~1138 at $z=2.2$ is a maturing protocluster (middle) and USS~1558 at 
$z=2.5$ is a growing protocluster (right), respectively, judging form substructures and 
characteristics of member galaxies. The left one represents the classical massive 
cluster of galaxies associated with a diffuse X-ray source. Galaxy colours are trying 
to express that the star formation declining and red sequence formation in the centre 
of galaxy clusters towards the local Universe. Yellow star symbols show energetic AGN 
activities. A table at the bottom highlights phenomena that are expected to occur in 
galaxies and (proto-)inter-cluster medium (see text).}
\label{fig13}
\end{figure*}

Combined with our previous paper \citep{Shimakawa:2018}, we here summarise the results 
that we obtained through the MDCS campaign. We have explored two protoclusters, 
USS~1558 ($z=2.53$) and PKS~1138 ($z=2.15$) with the deep narrow-band imaging on the 
Subaru Telescope ($NB_{3\sigma}\sim24$ mag). 

The former target, USS~1558 is thought to be a young protocluster composed of 
fragmented dense groups as seen at higher redshifts 
\citep{Umehata:2015,Oteo:2017,Miller:2018}. \citet{Shimakawa:2018} have found that 
SFRs of HAEs in dense group cores are statistically higher than those in inter-group 
regions. Also, local overdensities show the higher cut-off stellar mass in the stellar 
mass function than the inter-group. These trends suggest that galaxy formation is more 
enhanced and accelerated in the fragmented groups in a growing phase of hierarchical 
growth of galaxy clusters (see also \citealt{Wang:2016,Oteo:2017}). Such an active 
star formation could be supported by a rich cold gas reservoir (Tadaki et al. in 
preparation) fed by vigorous cold accretion (e.g., 
\citealt{Dekel:2006,Dekel:2009,Dekel:2009b,Keres:2009,Faucher:2010,Voort:2011}). 
Another companion paper \citep{Shimakawa:2017b} showing the \lya\ photon depletion in 
the dense group regions could support this scenario. 

PKS~1138 is thought to be a more advanced system because of the significant excess of 
bright red galaxies \citep{Kurk:2004a,Kodama:2007,Doherty:2010,Galametz:2012}. This 
work finds an enormous \ha\ structure associated with the Spiderweb radio galaxy and a 
high AGN fraction in massive HAEs. Such high energy sources could contribute 
pre-heating in the proto-inter-cluster medium \citep{Valentino:2016}, though heating 
mechanisms of inter-cluster medium in an early phase are highly uncertain 
observationally. Rest-frame $UVJ$ colours of X-ray HAEs imply that they would be in a 
post-starburst phase, suggesting the importance of the role of AGN as a driver of star 
formation quenching. These results, together with previous work 
\citep{Doherty:2010,Tanaka:2013,Koyama:2013b}, all point to PKS~1138 possibly being the 
sweet spot where maturing galaxies are undergoing a rapid transition from dusty 
starbursts to quenching populations. 

At lower redshifts ($z\lesssim2$), member galaxies in the hot inter-cluster medium 
enriched by superheated plasma would no longer retain their star formation because of 
insufficient gas accretion 
\citep{Keres:2005,Sijacki:2007,Schaye:2010,McCarthy:2011,Haines:2013,Hughes:2013,Jaffe:2015,Jaffe:2016,Bianconi:2016,Hayashi:2017}. 
Additionally, less massive galaxies falling into the clusters may experience 
environmental quenching as has previously been confirmed at $z<2$ (e.g., 
\citealt{Bamford:2009,Peng:2010,Smith:2012,Raichoor:2012,Muzzin:2012,Gobat:2015,Brown:2017}). 

Figure~\ref{fig13} briefly summarises these speculations based on the MDCS campaign. We 
strongly caution that the arguments discussed above are based on only two protoclusters 
and must therefore be viewed with caution. Future work is necessary to properly 
comprehend the systematic biases and appropriately understand the diversity of high-$z$ 
protoclusters, as mentioned in the following section.



\section{Future prospects: Diverse protoclusters}\label{s6}

Our MDCS program enables a high sampling density, down to a certain flux limit, with 
the deep narrow-band imaging: a unique advantage over the other common approaches such 
as photo-$z$ and the Balmer/Lyman Break technique. On the other hand, we targeted only 
two known protoclusters at $z>2$ which were initially discovered by surveys for radio 
galaxy environments \citep{Kurk:2000,Kajisawa:2006}. We thus should note that this 
work has not only a small sample size of protoclusters but also an unknown bias 
towards the environments associated with radio-loud galaxies. In particular, we do not 
yet know how the sampling effects affect the properties of member galaxies. 

In recent years, many intensive protocluster-search survey projects have been 
launched. These utilise continuous strong absorption in the \lya\ forest 
\citep{Cai:2016}, Lyman Break features \citep{Toshikawa:2017}, radio-loud galaxies 
\citep{Noirot:2018} and dust emission \citep{Planck:2015,Greenslade:2018} as 
signposts. Whilst these surveys will each construct statistical protocluster 
catalogues, the sampling effects remain to be resolved. Future work will be needed to 
understand such selection biases and categorise individual protoclusters with, e.g., 
X-ray properties \citep{Gobat:2011,Valentino:2016,Wang:2016} and \hi\ tomography 
\citep{Lee:2016}. The upcoming eROSITA \citep{Merloni:2012}, a successor to the 
ASTRO-H (Hitomi, \citealt{Takahashi:2016}), and the Prime Focus Spectrograph on the 
Subaru Telescope \citep{Tamura:2016} will possess the exceptional capability needed to 
reveal these unique objects.


\section*{Acknowledgements}

RS thanks Emma Rigby for proofreading the manuscript. 
We thank the anonymous referee for useful comments. 
The data are collected at the Subaru Telescope, which is operated by the National 
Astronomical Observatory of Japan, and also based on observations made with the 
NASA/ESA Hubble Space Telescope, and obtained from the Hubble Legacy Archive, which 
is a collaboration between the Space Telescope Science Institute (STScI/NASA), the 
Space Telescope European Coordinating Facility (ST-ECF/ESA) and the Canadian 
Astronomy Data Centre (CADC/NRC/CSA). We would like to thank the Subaru staff for 
their support throughout all observing and analysing processes. 
A part of analyses is conducted with the assistance of the Tool for OPerations on 
Catalogues And Tables ({\sc topcat}; \citealt{Taylor:2015}). This work gains the 
benefit from the 3D-HST Treasury Program (GO 12177 and 12328) with NASA/ESA HST, 
which is operated by the Association of Universities for Research in Astronomy, 
Inc., under NASA contract NAS5-26555. This research has made use of data obtained 
from the Chandra Source Catalogue, provided by the Chandra X-ray Centre (CXC) as 
part of the Chandra Data Archive, and also use of the NASA/ IPAC Infrared Science 
Archive, which is operated by the Jet Propulsion Laboratory, California Institute of 
Technology, under contract with the National Aeronautics and Space Administration.
R.S. acknowledges the support from the Japan Society for the Promotion of Science 
(JSPS) through JSPS overseas research fellowships. T.K. acknowledges KAKENHI No. 
21340045. HD acknowledges financial support from the Spanish Ministry of Economy and 
Competitiveness (MINECO) under the 2014 Ram\'on y Cajal program MINECO RYC-2014-15686. 
We wish to recognize and acknowledge the very significant cultural role and reverence 
that the summit of Maunakea has always had within the indigenous Hawaiian community. 
We are most fortunate to have the opportunity to conduct observations from this 
mountain.




\bibliographystyle{mnras}
\bibliography{bibtex_library} 


\appendix


\section{Colour-term effect}\label{a1}

We measured colour term variations ($K_s-NB_{2071}$) of individual galaxies, which are 
caused by the slight margin of filter wavelength between narrow-band ($NB_{2071}$) and 
broad-band ($K_s$) filters. Deriving the typical colour term is important not only to 
minimise the systematic error of derived narrow-band fluxes and EWs of HAEs but also 
to determine the reasonable EW threshold free of the contaminants from non-emitters. 

\begin{figure}
\centering
\includegraphics[width=0.9\columnwidth]{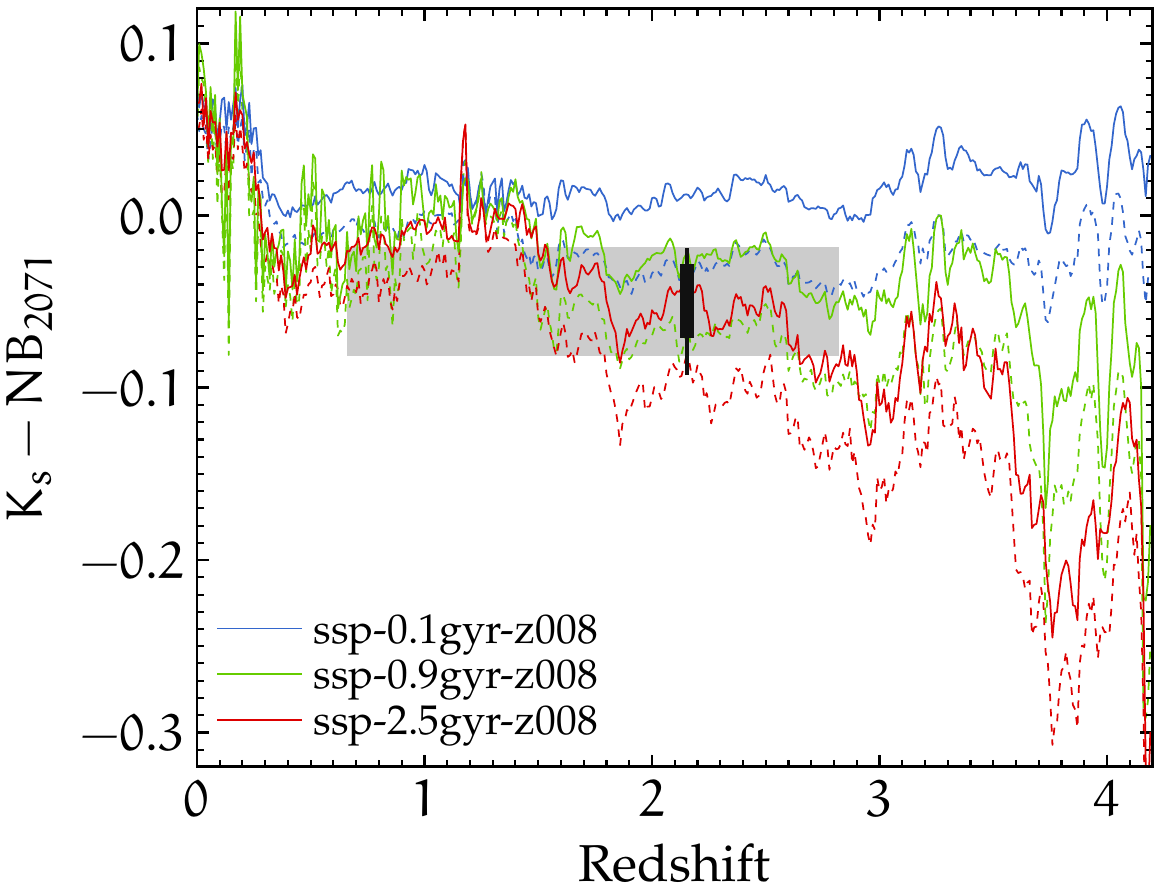}
\caption{Colour-term distribution of $K_s-NB_{2071}$ as a function of redshift in 
SSP models \citep{Bruzual:2003} with three different ages (from the top, 0.1 Gyr: 
{\it blue}, 0.9 Gyr: {\it green}, and 2.5 Gyr: {\it red}). Solid and dashed curves 
indicate zero extinction, and $A_V=1$, respectively. The metallicities are fixed to 
$Z=0.008$ in each model. The grey shaded region is the 68th percentile distribution 
inferred from SEDs of photo-$z$ sources at $K_s=$ 21--23 mag based on the 3D-HST 
photometry catalogue \citep{Skelton:2014}. Black thick and thin vertical lines 
indicate 68th and 95th percentiles of the colour-term distributions of HAEs estimated 
from the model-inferred SEDs with the {\sc fast} code (\S\ref{s2.4.1}).} 
\label{fig1a}
\end{figure}

We show three examples of the colour term distribution across the redshift range 
between 0 and 4 based on SEDs of single stellar population (SSP) synthesis models 
\citep{Bruzual:2003} with ages of 0.1, 0.9, and 2.5 Gyrs (fig.~\ref{fig1a}). We fix 
the stellar metallicity to $Z=0.008$ and apply two different extinctions ($A_V=0$ and 
1). It is easy to identify that the colour term decreases with increasing redshift, 
especially for older (i.e., redder) SED models. One should note that other models with 
different star formation histories or initial parameters do not significantly deviate 
from the range covered by these examples except when we employ extremely obscured 
(reddened) objects. 

Also, we demonstrated the colour terms of photo-$z$ sources from the 3D-HST library 
\citep{Skelton:2014} and obtained HAE samples whose SED spectra are based on the 
{\sc fast} SED-fitting code \citep{Kriek:2009}. We employ the same initial parameters 
in the SED-fitting for both samples (\S\ref{s2.4.1}). The inferred colour term 
variations can be found in fig.~\ref{fig1a}. We derived the colour term variation 
(68th percentile) of $-0.02\pm0.06$ in photo-$z$ sources at $K_s=$ 21--23 mag that is 
comparable with $K_s$-band magnitudes of most HAE samples (fig.~\ref{fig2}). This 
colour term variation almost corresponds to the 95th percentile distribution for the 
entire HAE samples. As mentioned in \S\ref{s2.3}, we decide to use the fixed colour 
term value of $K_s-NB_{2071}=-0.04$ since $<0.1$ mag scatter of the colour term 
effect is negligibly small relative to the total error in the narrow-band flux.


\section{Sample completeness}\label{a2}

\begin{figure}
\centering
\includegraphics[width=0.9\columnwidth]{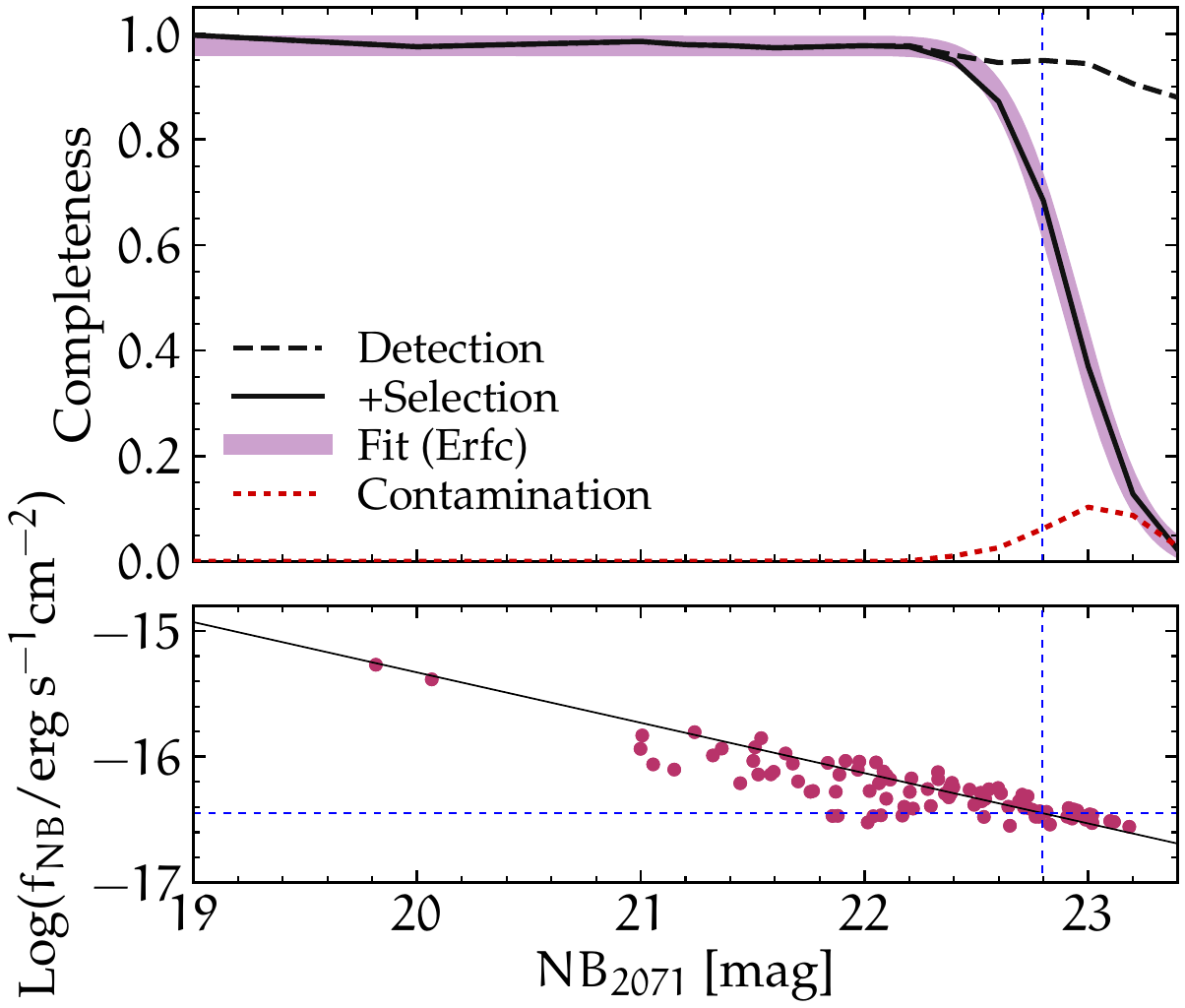}
\caption{Sample completeness as a function of narrow-band magnitude. Dashed and solid 
lines indicate the detection completeness and the completeness incorporating the 
selection process as well. We fitted the recovery rate by a complementary error 
function as shown by the thick purple curve. The red dotted line is the contamination 
rate. In the lower panel, red dots represent log narrowband fluxes ($f_{NB}$) across 
the narrow-band magnitude for the narrow-band emitters. In our simulation, we fix the 
embedded line flux into the narrow-band image by the narrow-band magnitude of PSFs 
along the regression line shown by the solid black line. The blue vertical and 
perpendicular lines indicate 68 percent completeness.}
\label{fig2a}
\end{figure}

Evaluating the completeness of narrow-band emitters requires an additional correction 
for the narrow-band selection process on top of the detection completeness. This work 
estimates the completeness as a function of narrow-band magnitude based on a Monte 
Carlo simulation in the same way as in \citet{Shimakawa:2018}. We first embedded 10 
PSF sources at the narrow-band magnitude of 19--25 mag, in steps of 0.2 mag, into the 
reduced $NB_{2071}$ and $K_s$-band images. The PSF model is made by median stacking 32 
bright point sources in the survey area. We should note that embedded PSFs at 
$NB_{2071}$ band implement line fluxes and the colour term dispersion in each PSF. 
Inserted narrow-band flux is determined by their narrow-band magnitude 
(fig.~\ref{fig2a}). Also, the colour term variation ($K_s-NB_{2071}=-0.02\pm0.06$) is 
derived from the result in the previous section. After that, we estimate the recovery 
rate in the detection process (i.e., detection completeness) and colour selection 
process (i.e., selection completeness) based on photometry with SExtractor (version 
2.19.5; \citealt{Bertin:1996}). In the latter case, we calculated the recovery rate 
for sources that satisfy the criteria for the narrow-band excess (eq.~\ref{eq1} and 
eq.~\ref{eq2}). This test was repeated 50 times for a given magnitude, and thus we 
simulated the total 500 PSF sources at each magnitude range. 

The obtained detection and selection completeness are indicated in fig.~\ref{fig2a}. 
We fit the resultant completeness fraction by a complementary error function 
(Erfc($x$)), 
\begin{equation}
C(NB) = 0.489 \times\mathrm{Erfc}(2.827\times(NB - 22.923)), \label{eq2a}
\end{equation}
which provides a better solution than polynomial and arctangent functions. 
$C(NB)^{-1}$ is the completeness correction factor that we applied to the derivation 
of the distribution function in \S\ref{s3.1}. According to our simulation, the 68 
percent completeness limit corresponds to a narrow-band magnitude of 22.80. One should 
note that narrow-band emitters with lower line fluxes should possess lower 
completeness for a given narrow-band magnitude. This factor could lead to a systematic 
error in the analyses of the distribution functions (\S\ref{s3.1}) at the faint end. 
We stress that such an error can be ignored as we remove faint-end sources for the 
curve fitting when we derive the distribution functions. 

This work also estimates the contamination rate caused by the combined effects of the 
photometric errors and the colour term. To do so, we run the same simulation without 
adding line flux to the narrow-band sources, and then calculate the contaminants 
showing flux excess at $NB_{2071}$ relative to $K_s$-band. We then confirm that our 
selection criteria (eq.~\ref{eq1} and \ref{eq2}) ignores such contaminations at 
$NB_{2071}<22.4$ mag. On the other hand, the simulation suggests a 5--10 percent 
contamination rate at the faint end. However, we do not consider this uncertainty 
since faint sources are outside the scope of this paper.


\section{Emitter catalogue}\label{a3}

The HAE catalogue is available as the online material. Descriptions 
about each column can be found in table~\ref{taba}. One should note that the 
SFR$_\mathrm{H\alpha,corr}$ error (eSFR\_Ha) would have additional uncertainties due 
to the systematic errors of dust extinction correction (see \S\ref{s3.3}), though the 
errors of model-inferred $A_V$ are included. 

\begin{table}
\centering
\caption{Catalogue columns.}
\begin{tabular}{ll}
\hline
Name & Description \\
\hline
ID & Identification number \\
RA & RA J2000 [degree] \\
Dec & DEC J2000 [degree] \\
$z$ & Spectroscopic redshift \\
$z$\_flag & $^\star$Spec-$z$ reference (id) \\
Ms & Log stellar mass [\msun] \\
Ms\_el & Lower $1\sigma$ error of Ms \\
Ms\_eu & Upper $1\sigma$ error of Ms \\
Av & Amount of dust extinction at $V$-band \\
Av\_el & Lower $1\sigma$ error of Av \\
Av\_eu & Upper $1\sigma$ error of Av \\
LHa & Observed \ha\ luminosity [erg~s$^{-1}$] \\
LHa\_e & $1\sigma$ error of $L_{H\alpha}$ [erg~s$^{-1}$] \\
SFR & Dust corrected log SFR$_\mathrm{H\alpha}$ [\msun~yr$^{-1}$] \\
SFR\_el & Lower $1\sigma$ error of SFR [\msun~yr$^{-1}$] \\
SFR\_eu & Upper $1\sigma$ error of SFR [\msun~yr$^{-1}$] \\
UV & Rest-frame ($U-V$) colour \\
VJ & Rest-frame ($V-J$) colour \\
X\_flag & X-ray counterpart (id in \citealt{Pentericci:2002}) \\
LAE & Ly$\alpha$ emitter counterpart (id in \citealt{Kurk:2004a}) \\
\hline
\multicolumn{2}{l}{$^\star$ P02: \citet{Pentericci:2002}, K04: \citet{Kurk:2004b}}\\
\multicolumn{2}{l}{C05: \citet{Croft:2005}, D10: \citet{Doherty:2010}}\\
\multicolumn{2}{l}{T13: \citet{Tanaka:2013}, S14: \citet{Shimakawa:2014}}\\
\multicolumn{2}{l}{Priority order: S14$>$T13$>$D10$>$C05$>$K04$>$P02}
\end{tabular}
\label{taba}
\end{table}


\section{The Spiderweb nebula}\label{a4}

\begin{figure}
\centering
\includegraphics[width=0.9\columnwidth]{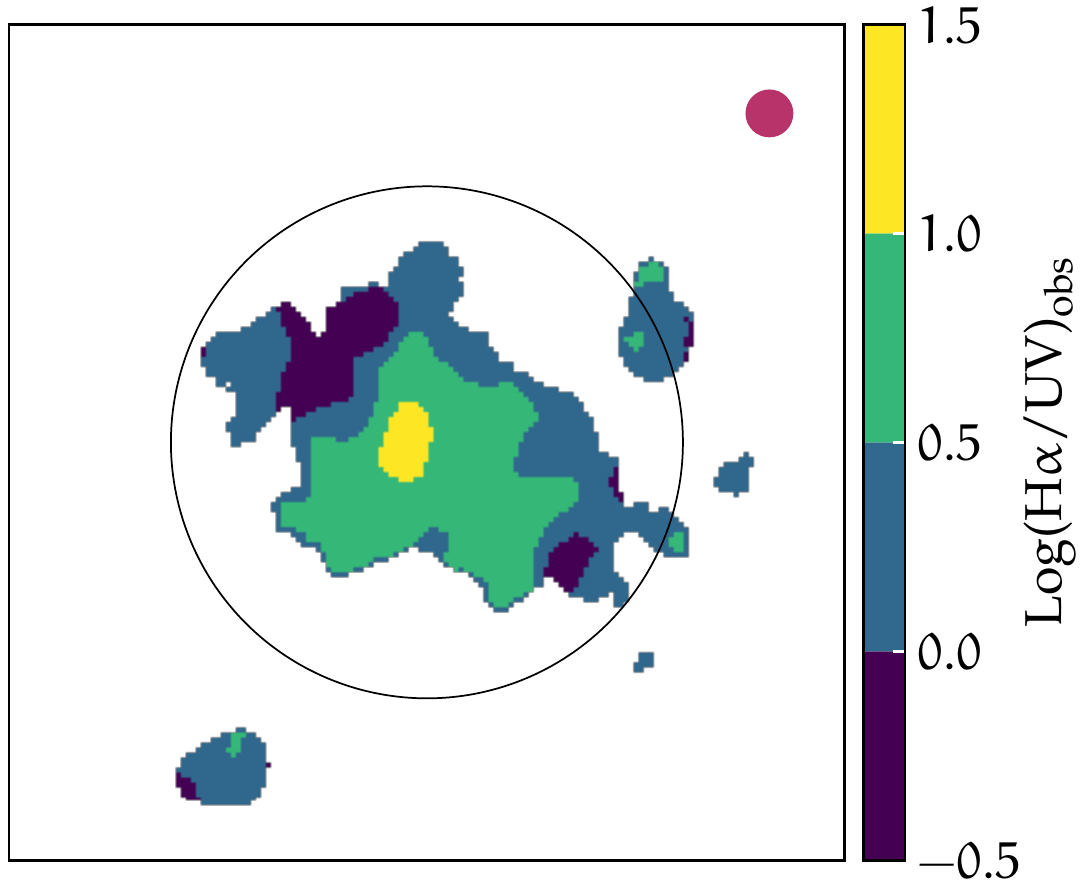}
\caption{Observed \ha/UV ratio map around the Spiderweb nebula. The colour bars on the 
right side shows the observed log \ha/UV ratio ($\equiv$ 
log(SFR$_\mathrm{H\alpha}$/SFR$_\mathrm{UV}$)). The SFRs are derived based on 1 
arcsec$^2$ area apertures as shown by the red circle. The figure only 
plots the regions with more than $2\sigma$ confidence levels in 
SB$_\mathrm{H\alpha+[N\textsc{ii}]}$. $1\sigma$ error of 
SB$_\mathrm{H\alpha+[N\textsc{ii}]}$ is $2.4\times10^{41}$ erg~s$^{-1}$arcsec$^{-2}$ 
for a given aperture area. We replace faint regions at F814W by $2\sigma$ limiting 
magnitude (27.28 mag in 1 arcsec$^2$ area). The black open circle indicates 50 ph-kpc 
distance from the centre.}
\label{fig3a}
\end{figure}

We compare the \ha\ intensity map with the UV surface brightness associated with the 
Spiderweb nebula to confirm if the extended \ha\ distribution needs additional \ha\ 
contributions from anything except star formation as examined by previous studies 
(e.g., \citealt{Kurk:2002,Nesvadba:2006}). We should note that these tentative analyses 
hold substantial error factors because of the filter flux loss and a lack of 3D 
spectral information. Figure~\ref{fig3a} shows the ratio map between \ha-based SFR 
(SFR$_\mathrm{H\alpha}$) and UV-inferred SFR (SFR$_\mathrm{UV}$). We derive the SFRs 
via the \citet{Kennicutt:1998} prescription and both measurements do not include 
extinction correction. The flux photometry is based on the 1 arcsec$^2$ 
aperture photometry by the SExtractor instead of deriving fluxes in each pixel to 
increase the S/N levels. We derive observed \ha\ flux from the 
narrow-band flux under the assumption that \nii/\ha\ line ratio is one 
\citep{Nesvadba:2006}. We estimate UV flux densities using the F814W image 
($\lambda_\mathrm{rest}\sim2500$ \AA\ at $z=2.15$). PSF size of the F814W image is 
matched to those of narrow-band and $K_s$-band (FWHM = 0.63 arcsec). We use $2\sigma$ 
limiting magnitude (F814W = 27.28 mag in 1 arcsec$^2$ area) in faint regions.

\begin{figure}
\centering
\includegraphics[width=0.9\columnwidth]{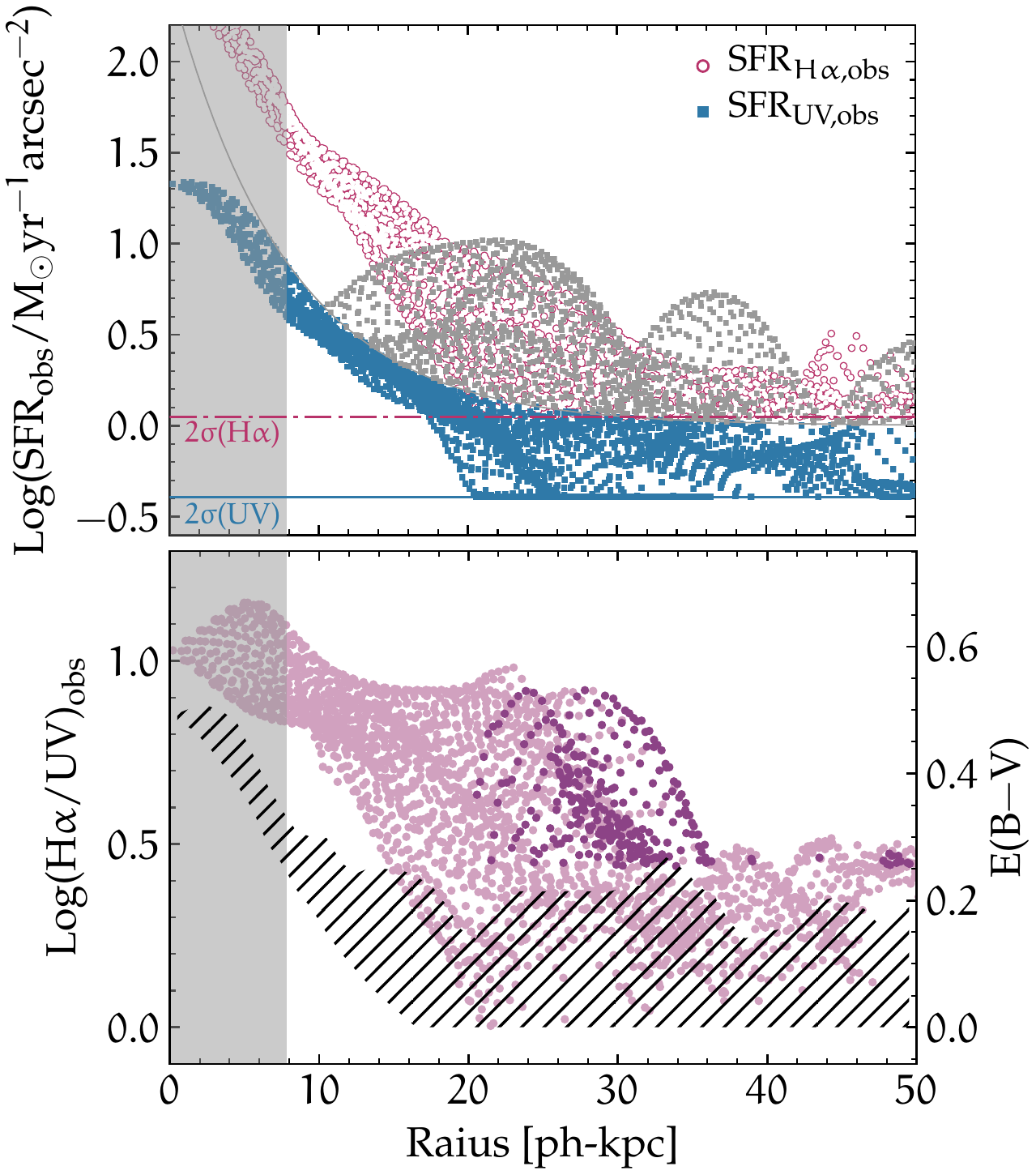}
\caption{{\it Upper panel:} Radial profile of observed SFR$_\mathrm{H\alpha,obs}$ (red 
open circles) and SFR$_\mathrm{UV,obs}$ (blue filled squares). We mask the UV-excess 
regions as shown by grey squares. The red dash-dotted line and the blue solid lines 
indicate $2\sigma$ limits in SFR$_\mathrm{H\alpha}$ and SFR$_\mathrm{UV}$ respectively. 
{\it Lower panel:} The observed \ha/UV ratio (purple circles). We employ $2\sigma$ 
limit mag in faint regions at F814W, which are highlighted by the darker colour. The 
black hatched areas show 95 percentile distribution of E(B$-$V) suggested by 
F475W$-$F814W colours assuming the \citet{Calzetti:2000} extinction law. The 
dust-reddening values correspond to the observed \ha/UV ratio shown in the left axis 
when we assume E(B$-$V)$_\mathrm{star}=$ E(B$-$V)$_\mathrm{nebula}$ and dust-corrected 
SFR$_\mathrm{H\alpha,corr}=$ SFR$_\mathrm{UV,corr}$.}
\label{fig4a}
\end{figure}

Radial distributions of \ha\ and UV -based SFRs are shown in fig~\ref{fig4a}. Each 
point corresponds to the aperture area centring on each pixel with a radial distance 
of 50 ph-kpc from the nucleus. To select diffuse \ha\ nebula regions and remove the 
bright UV components around the Spiderweb galaxy, we simply mask the regions showing 
excesses of UV-inferred SFRs, as shown in fig.~\ref{fig4a}. The figure thus only shows 
SFR$_\mathrm{H\alpha}$ at which UV-luminous regions have been filtered. The lower 
panel shows the radial profile of the \ha/UV ratio 
(SFR$_\mathrm{H\alpha}$/SFR$_\mathrm{UV}$) and corresponding dust-reddening values 
(E($B-V$)) needed to explain the excess of the ratio by assuming the 
\citet{Calzetti:2000} extinction law. We also plot the roughly-expected E($B-V$) 
suggested from the rest-UV colours between F475W and F814W assuming 
E($B-V$)$_\mathrm{star}=$ E($B-V$)$_\mathrm{nebula}$. 

Since a substantial uncertainty of the dust correction around the nucleus is expected, 
we do not discuss the central position. On the other hand, when we consider that the 
outer diffuse UV and \ha\ regions seem not to be heavily obscured as inferred from the 
$F475W-F814W$ colours, very high \ha/UV ratios at the 10--30 ph-kpc radii cannot be 
examined only by the dust reddening. This strong excess may require an extra \ha\ 
contribution from AGNs. These regions correspond to the south or west areas from the 
radio galaxy (fig.~\ref{fig3a}) where no detectably bright UV components can be found 
(fig.~\ref{fig12}). Interestingly, these spots correspond to the location showing the 
high \lya/\ha\ ratio (up to $\sim7.6$) and the radio jet \citep{Kurk:2002}. Our results 
seem to support their findings, although more detailed analyses are required to resolve 
the energy sources. 

This work does not compare in detail the radial distributions with those from other 
datasets such as the \lya\ and the CO($1-0$) images \citep{Kurk:2002,Emonts:2016} 
since these spatial resolutions are too large to resolve these UV luminous components. 
Future deep high-resolution observation would allow a more detailed comparison and 
provide new insights into the physical origins of this extended multi-phase gas 
nebula.

\bsp	
\label{lastpage}
\end{document}